\newcommand{\NPA}[3]{Nucl.\ Phys.\ A\ {\bf #1},\ #2 (#3)}
\newcommand{\NPB}[3]{Nucl.\ Phys.\ B\ {\bf #1},\ #2 (#3)}

\newcommand{\PLB}[3]{Phys.\ Lett.\ B\ {\bf #1},\ #2 (#3)}
\newcommand{\PR}[3]{Phys.\ Rep.\ {\bf #1},\ #2 (#3)}
\newcommand{\PRL}[3]{Phys.\ Rev.\ Lett.\ {\bf #1},\ #2 (#3)}

\newcommand{\PRC}[3]{Phys.\ Rev.\ C\ {\bf #1},\ #2 (#3)}
\newcommand{\PRD}[3]{Phys.\ Rev.\ D\ {\bf #1},\ #2 (#3)}

\newcommand{\EPJC}[3]{Eur.\ Phys.\ J.\ C\ {\bf #1},\ #2 (#3)}
\newcommand{\EPJA}[3]{Eur.\ Phys.\ J.\ A\ {\bf #1},\ #2 (#3)}

\newcommand\e{\epsilon}

\newcommand\q{\theta}

\newcommand\m{\mu}

\newcommand{\diracslash}[1]{#1\llap{/\kern2pt}}

\newcommand{\be}{\begin{equation}}
\newcommand{\ee}{\end{equation}}
\newcommand{\bea}{\begin{eqnarray}}
\newcommand{\eea}{\end{eqnarray}}
\newcommand{\ba}[1]{\begin{array}{#1}}
\newcommand{\ea}{\end{array}}

\documentclass[prd,aps,floats,nofootinbib,tightenlines,showpacs]{revtex4-1}
\usepackage{epsfig,graphicx,pstricks}
\usepackage{psfrag}
\usepackage{color}
\usepackage{amsmath}
\usepackage{amsfonts}
\usepackage{amssymb}
\usepackage{textcomp}
\usepackage{multirow}

\begin{document}

\title { Estimating transport coefficients in hot and dense quark matter}
\author{Paramita Deb }
\email{paramita@phy.iitb.ac.in}
\affiliation{Theory Division, Physical Research Laboratory,
Navrangpura, Ahmedabad 380 009, India}
\affiliation{Department of Physics, Indian Institute of Technology, Mumbai, India 400076}
\author{Guru Prakash Kadam}
\email{guru@theory.tifr.res.in}
\affiliation{Theory Division, Physical Research Laboratory,
Navrangpura, Ahmedabad 380 009, India}
\affiliation{Department of Theoretical Physics, Tata Institute of Fundamental Research, Homi Bhabha Road, Mumbai 400005, India}
\author{Hiranmaya Mishra}
\email{hm@prl.res.in}
\affiliation{Theory Division, Physical Research Laboratory,
Navrangpura, Ahmedabad 380 009, India}

\date{\today} 

\def\be{\begin{equation}}
\def\ee{\end{equation}}
\def\bearr{\begin{eqnarray}}
\def\eearr{\end{eqnarray}}
\def\zbf#1{{\bf {#1}}}
\def\bfm#1{\mbox{\boldmath $#1$}}
\def\hf{\frac{1}{2}}
\def\sl{\hspace{-0.15cm}/}
\def\omit#1{_{\!\rlap{$\scriptscriptstyle \backslash$}
{\scriptscriptstyle #1}}}
\def\vec#1{\mathchoice
        {\mbox{\boldmath $#1$}}
        {\mbox{\boldmath $#1$}}
        {\mbox{\boldmath $\scriptstyle #1$}}
        {\mbox{\boldmath $\scriptscriptstyle #1$}}
}

\begin{abstract}
We compute the transport coefficients, namely, the coefficients of shear and bulk viscosity as well as thermal
conductivity  for hot and dense quark
matter. The calculations are performed within the Nambu- Jona Lasinio (NJL) model.
The estimation of the transport coefficients is made using a quasiparticle approach of
solving the Boltzmann kinetic equation within the relaxation time approximation. The transition rates are calculated in a manifestly covariant
manner to estimate the thermal-averaged cross sections for quark-quark and quark-antiquark scattering. 
The calculations are performed for finite chemical potential also. Within the parameters of the model, the
ratio of shear viscosity to entropy density has a minimum at the Mott transition temperature. 
At vanishing chemical potential, the ratio of bulk viscosity to entropy
density, on the other hand, decreases with temperature with a sharp decrease near the critical temperature, and vanishes beyond it. At finite
chemical potential, however, it increases slowly with temperature beyond the 
Mott temperature. The coefficient of thermal conductivity also shows a minimum at the critical temperature.

\end{abstract}


\maketitle

\section{Introduction}
 Transport properties of hot and dense matter have attracted a lot of attention recently, in the context of relativistic heavy-ion 
collisions \cite{heinzrev}, as well as in astrophysical situations 
such as the early Universe \cite{berera} and perhaps
neutron stars \cite{pethick}. These properties enter as dissipative coefficients in the hydrodynamic evolution and therefore become 
an important ingredient for the near equilibrium evolution of the thermodynamic system. In the context of relativistic 
heavy-ion collisions, the matter produced at the core of the fireball 
is hot enough to have the quarks and gluons as the
the relevant degrees of freedom \cite{exptstar} that behaves like a strongly
 interacting liquid. This liquid with a small
shear viscosity expands, cools down, and, undergoes a phase transition
 to a hadronic phase to finally free stream to the detector.
One of the successful descriptions of this evolution is through 
relativistic hydrodynamics. A finite but very small
shear viscosity to entropy ratio ($\eta/s$) is necessary to explain the elliptic flow data\cite{hirano,schenke}.    The smallness of
$\eta/s$ is significant in connection with the conjectured 
lower bound $\eta/s=1/4\pi$, the 'Kovtun-Son-Starinets'(KSS) bound 
obtained in the context of AdS/CFT correspondence \cite{kss}.

The other viscous coefficient, the coefficient of bulk viscosity ($\zeta$), 
has recently been realized to be important to be included
in the dissipative hydrodynamics describing the evolution of QGP
 subsequent to a heavy-ion collision. This is because, the
bulk viscosity scales as the trace of the energy-momentum tensor and lattice
simulations indicate that the trace of energy-momentum tensor can be large near the transition temperature \cite{tanmoy}.
During the expansion of the QGP fireball, when the temperature approaches
the critical temperature $T_c$, the coefficient of  bulk viscosity
can be large and can give rise to different interesting 
phenomena like the phenomenon of cavitation 
when the pressure vanishes and the hydrodynamic
description of the evolution breaks down \cite{cavitation}. Indeed,
 there have been quite a few attempts to include the effects of
bulk viscosity on particle spectra and flow coefficients \cite{monai,kodama,schaferdus}. Effects of interplay of shear and bulk viscosity
on elliptic flow were investigated in \cite{ heinz,noronhahydro, rosegale} .

The transport coefficient that also plays an important role for
hydrodynamic evolution at finite chemical
potential is the thermal conductivity \cite{denicolhydro,denicolpre,denicolheat}. The effects 
of thermal conductivity in the relativistic hydrodynamics in 
the context of quark gluon plasma  have only recently
been studied\cite{rincon,denicolheat}. 

It is, therefore, desirable that these transport coefficients be
 understood and derived rigorously within a microscopic 
theory. They are not only interesting for their use in hydrodynamic 
simulations for the interpretation of
data generated in heavy-ion collision experiments, but also, 
in some cases, through their dependence on
the system parameters like temperature or chemical potential that
can  be indicative of a phase transition. For example, in many physical systems the coefficient of shear viscosity
is a minimum at the phase transition point. 
In principle, these transport coefficients can be estimated directly within QCD using the Kubo formulation \cite{kubo}. 
However, given that QCD is  strongly coupled for the energies accessible in  heavy-ion collision experiments, the task becomes complicated . Calculations with
first-principle lattice simulations at finite chemical potential is also challenging and  limited  only to the  equilibrium properties 
at small baryon chemical potential \cite{tanmoy,borsonyimu}. There have 
been numerous attempts to
estimate shear viscosity within various effective models \cite{sasakiqp,purnendu,ghosh,dobadoshear,prakashwiranata,greco,klevansky}
 involving different approximation schemes like relaxation time approximation, Chapman-Enskog and the Green Kubo formalism 
apart from weak coupling QCD \cite{transqcd}.  Most of these calculations have been performed with vanishing chemical potentials.
There have been attempts to estimate the transport coefficients using transport codes. The ratio, $\eta/s$, in the hadronic phase has been
estimated using UrQMD transport in Ref. \cite{bassprl} while the bulk and shear viscosity coefficients
have been estimated using parton hadron string dynamics(PHSD) transport code within a relaxation time approximation\cite{phsdbrat}.
The rise of bulk viscosity coefficient near the transition temperature has been observed in various effective models like
chiral perturbation theory \cite{dobadoch}, quasiparticle models \cite{quasip}, linear sigma model \cite{purnendu} and the Nambu-
Jona-Lasinio model \cite{sasakinjl}. Most of these calculations were performed at vanishing chemical potentials.
There have been attempts to compute bulk and shear viscosity at finite baryon densities. The ratio $\eta/s$ at finite $\mu$ 
was investigated using a relativistic Boltzmann equation for the pion nucleon system using a phenomenological scattering amplitude
\cite{chennakano,itakura}. Bulk viscosity at finite chemical potential has also been estimated with low energy theorems of QCD \cite{wang}.

The coefficient of thermal conductivity ($\lambda$) for strongly interacting matter has been estimated in different
 theoretical models. These include
 using kinetic theory for strongly interacting systems \cite{prakash}, chiral perturbation theory with a pion gas
\cite{nicola}, the Chapman-Enskog approximation \cite{matiello}; 
Green-Kubo approach within NJL model \cite{fukutome}
and the  instanton liquid model \cite{nam}. The results, however,
 vary over a  wide range of  values, with 
$\lambda=0.008$ GeV$^{-2}$ as in Ref. \cite{nicola} 
 to $\lambda \sim 10$ GeV$^{-2}$ as in Ref. \cite{marty} for a 
range of temperatures (0.12 GeV $<$T$<$ 0.17 GeV),  which  has 
been nicely tabulated in 
Ref. \cite{sabyath}. Thermal conductivity has also been calculated for vanishing baryon density but with a conserved pion number
in a pionic medium which can be relevant for heavy-ion collision system between kinetic and chemical freeze-out \cite{sourav,sabyath}.

 We shall here attempt to estimate these transport
coefficients within the NJL model. Estimations of the viscosity coefficients 
were made in Refs. \cite{klevansky,sasakinjl,marty} for the NJL model 
using a quasiparticle approach with a  Boltzmann kinetic equation.
We follow a similar approach of using the Boltzmann equation within a relaxation time approximation. We include here the
finite chemical potential effects \cite{sasakinjl} and also estimate the coefficient of thermal conductivity along with
the coefficients of shear and bulk viscosity. We might mention here
that both the medium dependence of the mass and the chemical potential bring out nontrivial contributions
to the  expressions for the viscosity coefficients, particularly, for the bulk viscosity. In Ref. \cite{voskresenskynpaa},
the authors discussed three different ansatze for the bulk
 viscosity expression 
in the quasiparticle approach when there are mean fields in the dynamical system and medium-dependent masses. This was
put on a firmer theoretical ground in Ref. \cite{albright}. 
The crucial observation made here was to realize that in the
ideal hydrodynamics, the entropy per baryon remains constant, which
restricts the variations of system parameters like temperature and chemical
potential while calculating first-order deviations.
The expressions for the viscosities turned out
to be a natural generalization of those at zero baryon density and are 
explicitly positive definite, as the dissipative
coefficients should be. We shall use here a similar approach to
derive the expressions for the transport coefficients within the NJL model.

We organize the present work as follows. In the following
section we discuss the two-flavor NJL model thermodynamics. We also recapitulate the medium
dependence of masses of the pion and sigma mesons here as these are needed to estimate the
scattering amplitudes of the quarks and antiquarks through meson exchange to estimate the
relaxation time. 
In Sec. III, we discuss the expressions for the shear and bulk viscosity within the
relaxation time approximation. In the next subsection we give the explicit calculations for the estimation of
transition rates at finite temperature and density  so as to estimate the medium-averaged
relaxation time. In Sec. IV, we give the
results for different transport coefficients.  Finally, in Sec. V we summarize our findings and give a possible outlook.

\section{Thermodynamics of two-flavor NJL model and  meson masses
}
We summarize here the thermodynamics of the simplest NJL model with two
flavors with a four-point interaction in the scalar and pseudoscalar channels,
with the Lagrangian given as
\be
{\cal L}=\bar\psi(i\gamma_\mu\partial^\mu-m_0)\psi-G\left((\bar\psi\psi)^2+
(\bar\psi i\gamma^5\zbf \tau^a\psi)^2\right).
\label{NJLL}
\ee
Here, $\psi$ is the doublet of $u$ and $d$ quarks. We also have assumed isospin
symmetry and have taken the same (current) mass $m_0$ for both flavors.
Using the standard methods of thermal field theory, one can write down the 
thermodynamic potential within a mean field approximation 
corresponding to the Lagrangian Eq.(\ref{NJLL})
as \cite{buballarev}
\bearr
\Omega(\beta,\mu)&=&-\frac{2N_cN_f}{(2\pi)^3}\int\sqrt{\zbf k^2+M^2} d\zbf k
\nonumber\\
&-&\frac{2N_cN_f}{(2\pi)^3\beta}\int d\zbf k\left(\ln(1+\exp(-\beta(E-\mu))+
\ln(1+\exp(-\beta(E+\mu))\right)\nonumber\\
&+&\frac{(M-m_0)^2}{4G},\nonumber\\
\label{pottot}
\eearr
where,  $\beta$ is the inverse of temperature, $\mu$ is the quark 
chemical potential, and,  $E(\zbf k)=\sqrt{\zbf k^2+M^2}$ is the on-shell
single-particle energy with `constituent' quark mass $M$. The 
constituent quark mass satisfies the self-consistent
gap equation
\be
M=m_0-2G\rho_s=m_0+\frac{2N_cN_f}{(2\pi)^3}\int\frac{M}{E(\zbf k)}
\left(1-f_-(\zbf k,\beta,\mu)-f_+(\zbf k,\beta,\mu)\right) d\zbf k,
\label{gapeq}
\ee
where, we have introduced the scalar density $\rho_s$,  given as
\be
\rho_s=\langle\bar\psi\psi\rangle=-\frac{2N_cN_f}{(2\pi)^3}
\int d\zbf k \frac{M}{E(\zbf k)}\left(1-f_-(\zbf k)-f_+(\zbf k)\right).
\label{rhos}
\ee
In the above $f_{\mp}(\zbf k,\beta,\mu)=\left(\exp(\beta(E\mp\mu))+1\right)^{-1}$ is the
fermion distribution function for quarks  and antiquarks, respectively, with
a constituent mass $M$ and,  are related to the quark number density
in the standard way:
\be
\rho=\frac{2N_cN_f}{(2\pi)^3}\int d\zbf k \left[f_-(\zbf k,\beta,\mu)
-f_+(\zbf k,\beta,\mu)\right] .
\ee
Within random phase approximation (RPA), the meson propagator can be calculated
as \cite{klevansky}
\be
D_M(\omega,\zbf p)=\frac{2iG}{1-2G\Pi_M(\omega,\zbf p)}
\label{rpaprop}
\ee
where, --$M=\sigma,\pi$ for scalar and pseudoscalar channel mesons, respectively,
and $\Pi_M$ is the polarization function in the corresponding mesonic channel.
The mass of the meson is extracted from the  pole position of the meson propagator
at zero momentum specified by the equation
\be
1-2G Re \Pi_M(m_M,\zbf 0)=0
\label{polemass}
\ee
Here, we have chosen to define the mass of the unbound resonance by the
real part of $\Pi_M$. For bound state solutions, i.e. for $\omega=m_M<2M$, the polarization function is
always real . For $m_M>2M$, $\Pi_M$ has an imaginary part that is related to the decay width  of the resonance as
$\Gamma_M=Im\Pi_M(m_M,\zbf 0)/m_M$.

Explicitly,
\be
\Pi_\pi(m_\pi,\zbf 0)=I_1-m_\pi^2 I_2(m_\pi,0)
\ee
\be
\Pi_\sigma(m_\sigma,\zbf 0)=I_1-(m_\sigma^2-2M^2) I_2(m_\sigma)
\ee

where,
\be
I_1=\frac{2N_cN_f}{(2\pi)^3}\int  \frac{d\zbf q}{E_q}
\left(1-f_-(\zbf q,\beta,\mu)-f_+(\zbf q,\beta,\mu)\right)
\label{i1}
\ee
and,
\be
I_2(m_{\pi/\sigma})=\frac{2N_cN_f}{(2\pi)^3}\int \frac{d\zbf q }{E_q}
\left(1-f_-(\zbf q,\beta,\mu)-f_+(\zbf q,\beta,\mu)\right)
\frac{1}{m_{\pi/\sigma}^2-4E(\zbf q)^2}
\label{i2}
\ee
so that the masses of the pion and sigma mesons are given, using the gap equation Eq.(\ref{gapeq}) by
\be
\frac{m_0}{M}+2Gm_\pi^2 Re I_2(m_\pi)=0
\label{pionmass}
\ee
for pions and
\be
\frac{m_0}{M}+2G(m_\sigma^2-4M^2)ReI_2(m_\sigma)=0
\label{sigmamass}
\ee
for  the mass of the sigma meson.
Explicitly, the real and the imaginary part of $\Pi_M(\omega,\zbf 0)$ are given as
\be
Re\Pi_M(\omega,\zbf 0)=\frac{2N_cN_f}{(2\pi)^3}
\int d\zbf q\frac{1}{E_{\zbf q}}\left(\frac{E_{\zbf q}^2-\epsilon_M/4}{E_{\zbf q}^2-\omega^2/4}
(1-f_-(E_{\zbf q}) -f_+(E_{\zbf q}))\right)
\ee
\be
Im\Pi_M(\omega,\zbf 0)=\theta(\omega^2-4 m^2)\frac{N_cN_f}{8\pi\omega}\left(\omega^2-\e_M^2\right)\left(1-f_-(\omega)-f_+(\omega)\right)
\ee
In the above, $f_\mp(x)=(1+\exp(\beta(x\mp\mu))^{-1}$ is the Fermi distribution function for particles and antiparticles.
It is easy to see that the meson propagators near the pole can be approximated by
 $D_M^{-1}(\sqrt s,\zbf 0)\sim \left(s-(m_M-i\Gamma_M/2)^2\right)$ with $m_M$ being the solution of Eq.(\ref{polemass}) and
and $\Gamma_M=Im\Pi_M/m_M$ \cite{zhuang}.  This will have interesting consequences while considering
quark scattering through exchange of mesons to estimate the relaxation time.

We might note here that, within the RPA approximation for the masses and widths of the
mesons as above,  the sigma meson has only a small width related to quark antiquark decay at low temperatures
and thus does not describe the scalar $f_0(500)$ meson which should have a
large pi-pi width that should be dominant for low temperature.
This is a limitation of the RPA method. In principle, one can go beyond the RPA to include mesonic fluctuations \cite{quack}
to include the low-temperature pionic width for the sigma meson. In what follows, however, we shall calculate the
transport coefficients due to quark scattering only  through meson exchange with the meson propagators
calculated within the RPA approximations in the NJL model, where
the elementary degrees of freedom are quarks. 

In the
following we look into the Boltzmann equation to derive the transport coefficients in terms of the 
relaxation time.

\section{ Boltzmann equation in relaxation time approximation and transport coefficients.}
 Within a quasiparticle approximation, a kinetic theory treatment for the calculation of transport
coefficient can be a reasonable approximation that we shall be following similar to 
that in Refs. \cite{sasakiqp,purnendu,voskresensky,blum,transqcd}. The plasma can be described by a phase space 
density for for each species of particle. Near equilibrium, the distribution function can be
expanded about a local equilibrium distribution
function for the quarks as,
$$
f(\zbf x,\zbf p,t) =f^0(\zbf x,\zbf p,t)+f^1(\zbf x,\zbf p, t)$$ where, the local equilibrium distribution function $f^0$ is given as
\be
f^0(\zbf x,\zbf p,t)=\left[\exp\left(\beta(x)\left(u_\nu(x)p^\nu\mp\mu(x)\right)\right)+1\right]^{-1}
\label{f0}
\ee
Here, $u_\mu=\gamma_u(1,\zbf u)$,  is the flow four-velocity, where,  $\gamma_u=(1-\zbf u^2)^{1/2}$;-- $\mu$ is the chemical
potential associated with a conserved charge like baryon number.
Further, $E_p=\sqrt{\zbf p^2+M^2}$ with a mass $M$ 
which in general is medium dependent. The departure from the equilibrium is 
described by the Boltzmann equation
\be
\frac{df_a}{dt}=\frac{\partial f_a}{\partial t}+\zbf v_a\cdot\zbf\nabla f_a-\zbf\nabla E_a\cdot\zbf\nabla_p f_a=-C^a[f]
\label{boltzeq}
\ee
where, --we have introduced the species index `$a$' on the distribution function.
With a medium-dependent mass, the last term on the left-hand side can be written as $(M/E_a)(\partial M/\partial x^i)(
\partial f^a/\partial p^i)$ and the Eq. (\ref{boltzeq}) can be recast as
\be
\frac{df_a}{dt}=\frac{p^\mu}{E_a}\partial_\mu f^a-\frac{M}{E^a}\frac{\partial M}{\partial x^i}\frac{\partial f^a}{\partial p^i}=
-C^a[f]
\label{boltz1}
\ee
To study the transport coefficients, one is interested in small departure from equilibrium in the hydrodynamic
limit of slow spatial and temporal variations. In the collision term on the right-hand side we shall be
limiting ourselves to $2\rightarrow 2$ scatterings only. Within the relaxation time approximation, in the
collision term for species $a$, all the distribution functions are given by the equilibrium distribution function except the 
distribution function for particle $a$. The collision term, to first order in the deviation from the equilibrium function,
will  then be proportional to $f_1$ , given the fact that $C^a[f_{0}]=0$ by local detailed balance.
In that case, the collision term is given by
\be
C[f]=-f_a^1/\tau_a,
\ee
 where, --$\tau^a$, the relaxation time for particle $`a'$, in general is a function of energy. However,  one can define 
a mean relaxation time by taking a thermal average of the
scattering cross sections which we shall spell out 
in the following subsection.

We shall next use the Boltzmann equation to calculate the transport coefficients in this relaxation time approximation. 
The departure from equilibrium
 for the distribution function is used to estimate the departure of the equilibrium energy-momentum tensor to define
the transport coefficients. Let us consider now the structure of the energy-momentum tensor $T^{\mu\nu}$ 
and of the quark current $J_\mu$. $T^{\mu\nu}$ and $J_\mu$ can be written in terms of chemical potential,
 temperature, and  four-velocity $u_\mu$ as
\be
T^{\mu\nu}=-pg^{\mu\nu}+w u^{\mu}u^{\nu}+\Delta T^{\mu\nu},
\label{tmunu}
\ee
and,-
\be
J_\mu=n u_\mu +\Delta J_\mu,
\label{jmu}
\ee
where, --$p(T,\mu)$ is the pressure, $\epsilon$ is the energy density, $w=\epsilon+p$ is the enthalpy, and $u_\mu$ is
the four-velocity of the fluid.  The dissipative parts are given by
\be
\Delta T^{\mu\nu}=\eta\left(D^\mu u^\nu+D^\nu u^\mu+\frac{2}{3}\Delta^{\mu\nu}\partial_\alpha u^\alpha\right)
-\zeta\partial_\alpha u^\alpha,
\label{deltatmunu}
\ee
and,
\be
\Delta J_\mu=\lambda\left(\frac{n T}{w}\right)^2 D_\mu\left(\frac{\mu}{T}\right)
\label{deltajmu}
\ee
where, --$\eta$, --$\zeta$ and $\lambda$ are the coefficients of shear viscosity, bulk viscosity, and thermal 
conductivity respectively. Further, in the above, $D_\mu=\partial _\mu-u_\mu u^\alpha\partial_\alpha$ is
the derivative normal to $u^\mu$. It is useful to note that, in the fluid rest frame, that will be used
to calculate the transport coefficients, $D_0=0$ and $D_i=\partial_i$. 

The energy-momentum tensor $T^{\mu\nu}$ and the current $J_\mu$ can also be written in terms of the distribution functions
as,--
\be
T^{\mu\nu}=\sum_a \int d\Gamma^a \frac{p^\m p^\nu}{E_a}f_a+g^{\mu\nu} V,
\label{tmunuf}
\ee
and,
\be
J_\mu=\int\sum_a t_a\int d \Gamma_a\frac{p_\mu}{E_a}f_a,
\ee
where, --we have introduced the notations $d \Gamma_a=g_a\frac{d^3p}{(2\pi)^3}$, $g_a$ being the degeneracy for species $a$--, and
$p^\mu=(E_a,\zbf p)$, with $E^a=\sqrt{\zbf p^2+m_a^2}$. Further, $V$ is the mean field or the ``vacuum" energy density contribution
in terms of the mean field giving a medium-dependent mass and $t^a=\pm 1 $ for particles and antiparticles respectively.
The nonequilibrium part of the distribution function is used to calculate the departure from equilibrium of the energy-momentum tensor.
The variation of the spatial part of Eq.(\ref{tmunuf}) is given as
\be
\delta T^{ij}=\sum_a\int d\Gamma^a\frac{p^i p^j}{E_a}
\left(\delta f_a-f_a^0\frac{\delta E_a}{E_a}\right) -\delta^{ij}\delta V,
\ee
where, the variation of the quasiparticle energy is also included to take into account the medium dependence of the mass.
The deviation of the distribution function, in general, will have departure from the equilibrium form. In addition it can
also change from the change in the single-particle energy from its equilibrium value. Defining the equilibrium values of $T$, $\mu$ and $E$
with a superscript $^0$, we can write
\be
\delta f_a=f_a(E_a,T,\mu)-f_a^0 (E_a^0,T^0,\mu^0)= \delta\tilde f_a-\frac{\delta E_a}{T}(f_a^0(1-f_a^0)),
\label{tildef}
\ee
where, we have defined $\delta \tilde f_a=f_a(E_a,T,\mu)-f_a(E_a,T_0,\mu_0)$ and have retained up to the linear term in $\delta E_a$.
Let us note that it is $\delta\tilde f_a$ that determines 
the transport coefficient as it is defined with 
the nonequilibrium energy, which enters in the
energy-momentum 
conservation in the collision term of the Boltzmann equation \cite{albright}.

Similarly, using the gap equation, the deviations in the vacuum energy term in Eq.(\ref{tmunuf}) is given by
\be
\delta V=\sum_a\int d \Gamma_a\frac{M}{E_a} f_a\delta M.
\ee
This leads to 
\be
\delta T^{ij}=\sum_a \int d \Gamma_a 
\frac{p^ip^j}{E^a}\delta\tilde f-\sum_a \int d \Gamma_a \frac{M}{E_a}f^a\left(1+\frac{p^2(1-f_a)}{3E_a T}
+\frac{\zbf p^2}{3E_a^2}\right)\delta M,
\label{deltatij}
\ee
where, --we have replaced $p^ip^j\sim 1/3(\zbf p^2)$ and for  the terms involving $\delta E_a$, we have used,$\delta E_a=(M/E_a) \delta M$.
 The terms involving $\delta M$ in Eq.(\ref{deltatij})
can be shown to vanish by doing a integration by parts leading to
\be
\Delta T^{ij}= \sum_a\int d \Gamma_a \frac{p^ip^j}{E_a}\delta\tilde f.
\label{dtij}
\ee

 In a similar manner, it can be shown that the departure of the quark current due to the nonequilibrium part of the distribution
function can be written as
\be
\Delta J^i=\sum_at_a\int d \Gamma^a\frac{\zbf p^i}{E_a}\delta\tilde f
\label{dji}
\ee
 Next, we compute  $\delta \tilde f_a\equiv f^1(x,p)$ using the Boltzmann equation, Eq.(\ref{boltz1}), in the relaxation time approximation.
This is then used to calculate nonequilibrium parts of energy-momentum tensor and the quark current to finally relate them to
 the transport equations using Eqs. (\ref{deltatmunu}) and (\ref{deltajmu}).
To do so, it is convenient to analyze in a local region choosing an appropriate rest frame. We further note that
we shall be working with first order hydrodynamics and hence will keep gradients up to
first-order in space time only. The left-hand side
of the Boltzmann equation Eq.(\ref{boltz1}), is explicitly small because of the gradients and we therefore may
replace $f^a$ by $f_0^a$.
In the local rest frame $u_\mu=(1,0,0,0)$, but, --the gradients of the velocities
are nonzero. Further, in the local equilibrium distribution function 
$f_a^0$ in Eq.(\ref{f0}), the flow velocity, temperature and chemical potential all depend upon $x$.
In addition, the four-momentum $p_a$ also depends upon the 
coordinate $x$ through the dependence of mass on the same.
We give here some details of the calculations of the left-hand side of the
Boltzmann equation. To do so, let us calculate  the derivative of
the equilibrium distribution function Eq.(\ref{f0}) given as
\be
{\partial_\mu f_a^0}=-f_a^0(1-f_a^0)\left[-\frac{1}{T^2}(E_a-\mu_a)+\frac{1}{T}
\partial_\mu(p_\nu u^\nu-\mu^a)\right]
\ee
Noting the fact the $E^a$ also has spatial dependence through is mass
dependence, one obtains  for the first term in the L.H.S. of Eq.(\ref{boltz1})
\be
\frac{p^\mu}{E^a}\partial_{\mu} f_a^0=\frac{f_a^0(1-f_a^0)}{E^a}\left[
\frac {E^a}{T^2}p^\mu\partial_\mu T
+p^\mu\partial_\mu\left(\frac{\mu^a}{T}\right)
-\frac{1}{T}\left(p^\mu\partial_\mu E^a+p^\mu p^\nu\partial_\mu u_\nu\right)
\right]
\label{term1}
\ee
while, the second term is given as
\be
\frac{\partial f_a^0}{\partial p^i}=-f_a^0(1-f_a^0)\frac{p_i}{E^aT}
\label{term2}
\ee
Next using the fact that $u^\nu u_\nu=1$, one can show that,
 in the local rest frame,
 $\partial_\nu u^0=0$. This can be used to expand the term with gradient of flow velocity
 Eq.(\ref{term1}) in terms of spatial and 
temporal derivatives of the flow velocity $u_i$.
Combining both Eq.(\ref{term1}) and Eq.(\ref{term2}),
LHS of Eq.(\ref{boltz1}), is given as,
\be
\frac{f_a^0(1-f_a^0)}{E^a}\left[-E^a\partial_0
\left(\frac{E^a-\mu^a}{T}\right)
\frac{E^a p^i}{T}\left(\frac{\partial_i T}{T}-\partial_0u_i\right)
+p^i\partial_i\left(\frac{\mu^a}{T}\right)-p^ip^j\partial_ju_i\right]=
-\frac{f_a^1}{\tau}.
\ee
Next, we can use the conservation equation $\partial_\mu T^{\mu\nu}=0$ to
write $\partial_0u_i=\partial_i P$ in the rest frame .
One can use  thermodynamic relations $\partial_i P=s\partial_iT+n\partial_i\mu$
to write $(\partial_i T)/T-\partial_0 u_i=-(nT/w)\partial_i(\mu/T)$. Further, 
the spatial derivative of the flow velocity can be decomposed in to
a traceless part and a divergence part in the flow velocity. 

This leads to
\be
\frac{df_a^0}{dt}=\frac{f_a^0(1-f_a^0)}{T} q^a(\beta,\mu)=-\frac{\delta\tilde f_a}{\tau_a}
\label{boltz2}
\ee
 where,we have defined,
\bearr
q^a(T,\mu)&=&-\bigg[
\frac{\partial T}{\partial t}\left(\frac {E^a-\mu^a}{T}-\frac{\partial E^a}{\partial T}\right)\nonumber\\
&-&\frac{\partial \mu}{\partial t}\left(\frac{\partial E^a}{\partial\mu}-t^a\right)+\frac{T}{E^a}\left(t^a-\frac{E^an}{w}\right)
p^i \partial_i\left(\frac{\mu}{T}\right)\nonumber\\
&-&\frac{p^ip^j}{2E^a}W_{ij}+\frac{\zbf p^2}{3E^a}\partial_ku^k\bigg]\nonumber\\
\label{qa}
\eearr
  The Boltzmann equation  Eq.(\ref{boltz2}) thus relates the non equilibrium part of the distribution functions to the
 variation in fluid velocity,the temperature and the chemical potential. This will be used to calculate the dissipative part of the 
energy momentum tensor. 

Using stress-energy conservation $\partial_\mu T^{\mu\nu}=0$; --the baryon number conservation equation $\partial_\mu J^{\mu}= 0$,
 and  standard thermodynamic relations, one can relate the  temporal derivatives of temperature and chemical
potentials with the velocity of sound at constant baryon density and constant entropy density,  respectively,
 as
\be
\partial_0T=-v_n^2T\zbf\nabla\cdot\zbf u
\label{tdot}
\ee
and 
\be
\partial_0\mu=-v_s^2\mu\zbf\nabla\cdot\zbf u.
\label{mudot}
\ee
The velocity of sound at constant density($n$) or at constant entropy ($s$)  can be calculated using Jacobian methods
as
\be
v_n^2=\left(\frac{\partial P}{\partial \epsilon}\right)_n=\frac{\partial (p,n)}{\partial(\epsilon,n)}=\frac{s\chi_{\mu\mu}-n\chi_{\mu T}}{
\frac{\partial\epsilon}{\partial T} \chi_{\mu\mu} -\frac{\partial\epsilon}{\partial\mu}\chi_{\mu T}}
\label{vn2}
\ee
and
\be
v_s^2=\left(\frac{\partial P}{\partial \epsilon}\right)_s=\frac{\partial (p,s)}{\partial(\epsilon,s)}=\frac{s\chi_{\mu T}-n\chi_{T T}}{
\frac{\partial\epsilon}{\partial T} \chi_{\mu T} -\frac{\partial\epsilon}{\partial\mu}\chi_{T T}}
\label{vs2}
\ee
In the NJL model one can explicitly calculate the derivatives of the energy density with temperature or chemical potential. 
On the other hand, using thermodynamic relations one can also rewrite Eqs.(\ref{vn2})  and (\ref{vs2}) as
\be
v_n^2=\frac{s\chi_{\mu\mu}-n\chi_{\mu T}}{
T( \chi_{\mu\mu}\chi_{TT} -\chi_{\mu T}^2)}
\label{vn22}
\ee
\be
v_s^2=\frac{n\chi_{TT}-s\chi_{\mu T}}{
\mu( \chi_{\mu\mu}\chi_{TT} -\chi_{\mu T}^2)}
\label{vs22}
\ee
Thus, we can have from Eqs.(\ref{tdot}) and (\ref{mudot}), the variation for the distribution function in the relaxation time approximation,
\be
\frac{\delta \tilde f_a}{\tau_a}=-\frac{f_a^0(1-f_a^0)}{T}q^a(T,\mu),
\label{deltaftilde}
\ee
with $q^a(T,\mu)$ given as
\be
q^a(T,\mu)=-Q_a(T,\mu, \zbf p^2)\zbf\nabla\cdot\zbf u+\frac{T}{E_a}
p^i\partial_i\left(\frac{\mu}{T}\right)\left(t_a-\frac{E_an}{w}\right)
+\frac{p^ip^j}{2T}W_{ij}
\label{qqa}
\ee
In the above, the coefficient of the divergence in flow velocity part, $Q_a$, is given by
\be
-Q_a(T,\mu,\zbf p^2)=\bigg[v_n^2\left(-E_a+T\frac{\partial E_a}{\partial T}+\mu\frac{\partial E_a}{\partial \mu}\right)
+\left(\frac{\partial P}{\partial n}
\right)_\epsilon\left(\frac{\partial E}{\partial\mu}-t^a\right)
+\frac{\zbf p_a^2}{3E_a}\bigg]
\label{Qa}
\ee

 Substituting the expression for $\delta \tilde f$ from Eq.(\ref{deltaftilde}) in Eq.(\ref{dtij}), in the local rest frame,
\be
\delta T^{ij}=\sum_a\int d\Gamma \frac{p_a^ip_a^j}{TE_a}\tau_a f_a^0(1-f_a^0)
q_a(\zbf p,\beta,\mu).
\label{deltij}
\ee
The contribution of the term proportional to the gradient 
of the $(\mu/T)$ term in Eq.(\ref{qqa}) vanishes because of
symmetry. When comparing the resulting expression with the tensor structure of the dissipative part of $\Delta T^{\mu\nu}$
of Eq.(\ref{dtij}), we have the expressions for the shear viscosity coefficient $\eta$ as
\be
\eta=\frac{1}{15 T}\sum_a\int d\Gamma_a\frac{\zbf p_a^4}{E_a^2}
\left(\tau_af_a^0(1-f_a^0)) \right).
\label{defeta}
\ee
Similarly, the bulk viscosity coefficient $\zeta$ is given as
\be
\zeta=-\frac{1}{3T}\sum_a\int d\Gamma_a\frac{\zbf p_a^2}{E_a}\left(\tau_af_a^0(1-f_a^0)Q_a\right)
\label{defzeta}
\ee

In a similar manner, one can substitute $\delta \tilde f$ in Eq.(\ref{dji})to obtain
\be
\Delta J_i=\sum_a\int d\Gamma_a p^i\tau^a f_a^0(1-f_a^0)q^a(t,\mu).
\label{ddji}
\ee
In the above, in contrast to Eq.(\ref{deltij}), the term in $q^a$ that results in a nonzero contribution from $\q^a(t,\mu)$,
is the term with gradient in $(\mu/T)$. Comparing this with Eq.(\ref{deltajmu}), we have the thermal conductivity given as
\be
\lambda=\left(\frac{w}{nT}\right)^2\sum_a\int d\Gamma^a\frac{\zbf p^2\tau_a}{3E_a^2}\left(1-\frac{t^anE^a}{w}\right)
 f_a^0(1-f_a^0) 
\label{lambdatau}
\ee

However, the solutions for $Q^a$ as given in Eq.(\ref{Qa}) for the bulk viscosity is
to be supplemented by Landau-Lifshitz matching conditions, i.e., the variations of the 
distribution function should be such that they satisfy
the conditions $u_\mu\Delta J^\mu=0$ 
and $u_\mu\Delta T^{\mu\nu}u_\nu=0$. In the local rest frame
these conditions reduce to
\be
\Delta J_0=\sum_a\int d\Gamma_a t^a\delta f_a=0
\label{ll0}
\ee
\be
\Delta T^{00}=\sum_a\int d\Gamma_a E_a\delta f_a=0.
\label{ll00}
\ee
Using Eq.(\ref{tildef}) relating $\delta f_a$ and $\delta \tilde f_a$, one can write
the Landau-Lifshitz conditions in the relaxation time approximation as
\be
\Delta J_0=\langle \tau^a Q^a(T,\mu)t^a g^a(T,\mu)\rangle =0
\label{ll1}
\ee
\be
\Delta T_{00}=\langle \tau^a Q^a(T,\mu)E^a g^a(T,\mu)\rangle =0
\label{ll2}
\ee

with,
\be
g^a(T,\mu)=1-\frac{T\left(\frac{\partial E^a}{\partial T}\right)_\sigma}{E^a-\mu^a+T\left(\frac{\partial\mu^a}{\partial T}\right)_\sigma}
\label{ga}
\ee
where, we have defined the derivative with respect to temperature at fixed entropy per quark as \cite{albright}
\be
\left(\frac{\partial E^a}{\partial T}\right)_\sigma=\left(\frac{\partial E^a}{\partial T}\right)_\mu+
\left(\frac{\partial E^a}{\partial \mu}\right)_T\left(\frac{\partial \mu}{\partial T}\right)_\sigma
\ee
and
\be
\left(\frac{\partial \mu}{\partial T}\right)_\sigma=\frac{1}{T}\left[\mu+\frac{1}{v_n^2}\left(\frac{\partial p}{\partial n}\right)_\epsilon
\right]
.
\label{delmudelt}
\ee
The above arises due to the fact that the variations of temperature and chemical potential are not independent variations. They
 are related by
the hydrodynamic flow of the matter which occurs at constant entropy per baryon $\sigma=s/n$ \cite{albright}. 
Further, we have introduced the notation\cite{voskresenskynpaa}
$$\langle\phi_a(p)\rangle=\int d\Gamma_a[\phi_a(p)f_a^0(1-f_a^0)].$$

If the variations as in Eq.(\ref{deltaftilde}) do not satisfy the the Landau-Lifshitz conditions Eqs.(\ref{ll1}) and (\ref{ll2}),
one may still fulfill them by performing a shift\cite{voskresenskynpaa,albright}
\be
\tau_aQ_a\rightarrow \tau_aQ_a-\alpha_n t^a-\alpha_e E^a
\label{replace}
\ee
where, $\alpha_n$ and $\alpha_e$ are the Lagrange multipliers   associated with conservation of baryon number and energy. 
Performing the substitution Eq.(\ref{replace}) in Eqs. (\ref{ll1}) and (\ref{ll2}) we have the Landau-Lifshitz conditions given as

\be
\sum_at_a\langle\tau_a Q_a\rangle-\alpha_n\sum_a\langle g^a\rangle-\alpha_e
\sum_a \langle t^aE^a g^a\rangle=0,
\label{ll}
\ee
\be
\sum_a\langle E^a\tau^a Q^a\rangle-\alpha_n\sum_a\langle t^aE^ag^a\rangle-\alpha_e
\sum_a \langle E_a^2 g^a\rangle=0.
\label{llp}
\ee
One can solve these two equations for the coefficients $\alpha_e$ and $\alpha_n$ and calculate the bulk viscosity coefficient
$\zeta$ after performing the replacement Eq.(\ref{replace}) in Eq.(\ref{defzeta}).
This leads to 
\be
\zeta=-\frac{1}{3T}\sum_a\int d\Gamma_a\frac{\zbf p_a^2}{E_a}\left(\tau_af_a^0(1-f_a^0)Q_a\right)-\alpha_e w-\alpha_n n.
\label{defzeta1}
\ee
On the other hand, it is convenient to use Eqs.(\ref{ll}) and (\ref{llp}) to obtain
\be
\alpha_e w+\alpha_n n=-\sum_a\langle \tau^a Q^a\left(E^a-T\frac{\partial E^a}{\partial T}-\mu\frac{\partial E^a}{\partial \mu }\right)
+\left(\frac{\partial P}{\partial n}\right)_\epsilon\left (\frac{\partial E^a}{\partial \mu}-t^a\right )\rangle.
\ee
Substituting this back in Eq.(\ref{defzeta1}), we have
\bearr
\zeta&=&\frac{1}{9 T}\sum_a\int d\Gamma^a\tau^a f_a^0(1-f_a^0)\nonumber\\
&\times&
\left[\frac{\zbf p^2}{E^a}-3v_n^2\left(E^a-T\frac{\partial E^a}{\partial T}-\mu
\frac{\partial E^a}{\partial\mu}\right)+
3\left(\frac{\partial P}{\partial n}\right)_\epsilon
\left(\frac{\partial E^a}{\partial\mu}-t^a\right)\right]^2
\nonumber\\
\label{zeta2}
\eearr

In a similar manner, putting the constraint $\Delta T^{0i}=0$ in the rest frame yields, the expression for
thermal conductivity as\cite{albright}
\be
\lambda=\frac{1}{3}\left(\frac{w}{nT}\right)^2\sum_a\int d\Gamma\frac{\zbf p^2}{E_a^2}\tau^a\left(t^a-\frac{nE^a}{w}\right)^2
f_a^0(1-f_a^0)
\label{lambda2}
\ee 
In passing, we would like to comment here that the expression for thermal conductivity is identical to those as derived in 
Refs. \cite{gavin,hosoya}.

Thus all the dissipative coefficients are explicitly positive definite within the relaxation time approximation. The expression
for the bulk viscosity reduces to the expression for the same in the limit of vanishing density to that of Ref.\cite{purnendu}.
Further, the expression also reduces to the expression for bulk viscosity in Ref. \cite{gavin} when the medium dependence 
of the single-particle energy is not taken into account.  We would like to comment here that, the difference between including the
Landau-Lifshitz condition Eqs.(\ref{ll0}) and (\ref{ll00}), and not including the same has been pointed out in Ref.\cite{marty}
for $\mu=0$. 
 Equations (\ref{defeta}), (\ref{zeta2}) and (\ref{lambda2}) for the dissipation coefficients shall be the 
focus of our  discussion in what follows. Let us note that in these equations so far, the unknown quantity is the 
estimation of the relaxation time $\tau^a$.  As mentioned earlier, $\tau^a$, in general,  will be energy dependent but
we shall be taking an energy-averaged estimation of the relaxation time by taking the thermal average of the scattering cross section.

\subsection{Transition rates and thermal averaging}

 The key quantity in estimating the transport coefficient 
is the thermal-averaged transition rate $\bar W$ to estimate the average relaxation time $\tau$. This has been dealt with e.g. 
in Refs. \cite{sasakinjl,marty} by multiplying the zero-temperature cross section with the Pauli blocking factor and then taking an energy average weighted by a normalized
distribution function to calculate the mean cross section and hence the relaxation time. On the other hand, we follow a
procedure of thermal averaging in a manner similar to Ref.\cite{gondolo} which is manifestly Lorentz covariant. Such an averaging procedure has
been performed in ref.\cite{klevanskynpa}. The difference between the two approaches has also been discussed in Ref. \cite{klevanskynpa}.
The average transition rate $\bar W$, e.g., for a general fermion-fermion scattering process $a,b\rightarrow c,d$
is given as
\be
\bar W_{ab}=\frac{1}{n_a n_b}\int d\pi_a d\pi_b f^a(p_a)f^b(p_b) W_{ab}(s)
\label{wbarab}
\ee
In the above, $f_i$ are the distribution functions for the fermions  and $d\pi_i=(1/(2\pi)^3) d\zbf p^i/2E_i$, $n_i=(g_i/(2\pi)^3)\int
d\zbf p_i f(\zbf p_i) $ is the number density of i-th species with degeneracy $g_i$. Further, the quantity 
$W_{ab}(s)$  is dimensionless,
Lorentz invariant and is dependent only on the Mandelstam variable $s$, and it is given as
\be
W_{ab}(s)=\frac{1}{1+\delta_{ab}}
\int d\pi_c d\pi_d (2\pi)^4\delta^4(p_a+p_b-p_c-p_d)|\bar M|^2(1-f_c(\zbf p_c))(1-f_d(\zbf p_d)).
\label{wabs}
\ee

Here, we have included the Pauli blocking factors.  The quantity $W_{ab}(s)$ can be related to the cross section by noting that
\be
\frac{d\sigma}{dt}=\frac{1}{64\pi s} \frac{1}{p^2_{ab}} |\bar M|^2
\ee
 where $p_{ab}=\sqrt{(s-4m^2)}/2$ is the magnitude of the three momentum of the incoming
particles in the center-of-mass (CM) frame if the masses of the particles are the same. Thus in the CM frame,we have, using the delta
function and  integrating over the final momenta
\be
W_{ab}(s)=\frac{2\sqrt{s(s-4m^2)}}{1+\delta_{ab}}\int_{t_{min}}^{0}dt\left(\frac{d\sigma}{dt}\right)
\left(1-f_c(\frac{\sqrt s}{2},\mu)\right)\left(1-f_d(\frac{\sqrt s}{2},\mu)\right)
\label{wsig}
\ee
where,  $t_{min}=-(s-4m^2)$ for the nonidentical particle case and $t_{min}=-1/2(s-4m^2)$ for the case of  identical
particles in the final state.

Once $W_{ab}$ is calculated as a function of $s$, one has to do the thermal averaging
of the transition rate using Eq.(\ref{wbarab}). To perform the integration over $d\pi_a d\pi_b$ in Eq.(\ref{wbarab}),
we note that the volume element $d\zbf p_a d\zbf p_b$ is given by
\be
d\zbf p_a d\zbf p_b=4\pi|\zbf p_a|E_a d|\zbf p_a| 4\pi|\zbf p_b|E_b d|\zbf p_b| \frac{1}{2}d(\cos\theta),
\ee
where, $\theta$ is the angle between the three-momenta $\zbf p_a$ and $\zbf p_b$
It is somewhat convenient to change the integration variables from $E_a, E_b,\theta$ to $E_+, E_-, s$ given by
$$E_+=E_a+E_b,\quad E_-=E_a-E_b$$ $$s=2m^2+2E_aE_b-2|\zbf p_a||\zbf p_b|\cos\theta$$
so that the volume element becomes
\be
d\zbf p_a d\zbf p_b=2\pi^2E_aE_bdE_+dE_-ds
\ee

The integration region $(E_1>m,E_2> m ,|\cos\theta|\le 1)$ transforms into
$$|E_-|<X,
\quad E_+\ge \sqrt s , \quad s\ge 4 m^2$$, 
\noindent where, $X= \sqrt{1-\frac{4m^2}{s}}\sqrt{E_+^2-s}$.  It is then possible to perform the integration over the variable $E_-$
analytically when the distribution functions in Eq.(\ref{wbarab}) are fermionic distribution functions
 $f(x)=(1+\exp(\beta (x-\mu)))^{-1}$. 
Thus the thermal-averaged transition rate is given by
\be
\bar W_{ab}=\frac{1}{n_a n_b} \frac{g_a g_b}{(2\pi)^4}\frac{1}{8}\int_{4m^2}^{\infty} ds\int_{\sqrt s}^{\infty}dE_+\int_{-X}^{X}
dE_- f^a(\frac{E_a+E_b}{2},\mu,\beta)f^b((E_a-E_b),\mu,\beta) W_{ab}(s)
\label{barw}
\ee
The thermal relaxation time for each species is then given as \cite{degroot}
\be
\tau^{-1}_a=\sum_bn_b\bar W_{ab}\equiv\bar\omega_a=\frac{1}{n_a}\sum_b\int\frac{d\zbf p_a}{(2\pi)^3}\omega_a(E_a)f(E_a)
\label{tauinv}
\ee
where, --we have defined a mean interaction frequency $\bar\omega_a$ similar to Ref.\cite{purnendu}
with the energy-dependent interaction frequency given as
\be
\omega_a(E_a)=\frac{1}{2E_a}\int d\pi_b f(E_b) W_{ab}.
\label{omgea}
\ee

 In Eq.(\ref{tauinv}), the summation runs over all species of quarks and $\bar W_{ab}$ is the sum of the transition rates 
of all processes with $a$,
$b$ as the initial states. 
In the present case of two flavors we consider the following possible scattering. 
$$u\bar u\rightarrow u\bar u,\quad u\bar d\rightarrow u\bar d,\quad u\bar u\rightarrow d\bar d,$$
$$u u\rightarrow u u,\quad u d\rightarrow u d,\quad \bar u\bar u\rightarrow\bar u \bar u,$$
$$\bar u\bar d\rightarrow \bar u\bar d,\quad d\bar d\rightarrow d\bar d,\quad d\bar d\rightarrow u\bar u,$$
$$d\bar u\rightarrow d\bar u,\quad d d\rightarrow d d,\quad \bar d\bar d\rightarrow \bar d\bar d,$$
One can use $i$-spin symmetry, charge conjugation symmetry and crossing symmetry to relate the matrix element
square for the above 12 processes to get them related to one another and one has to evaluate only two independent matrix elements
to evaluate all the 12 processes. We can choose these, as in Ref. \cite{klevansky}, to be the processes
$u\bar u\rightarrow u\bar u$ and $u\bar d\rightarrow u\bar d$ and use the symmetry conditions to calculate the rest. We note however
that while the matrix elements are related, the thermal-averaged rates are not,  as they involve also the thermal 
distribution functions for the initial states as well as the Pauli blocking factors for the final states.  For the sake of
completeness we also write down the square of the  matrix elements for these two processes explicitly which is given in Ref.\cite{klevansky}.
This
for the process $u\bar u\rightarrow u\bar u$ is given as \cite{klevansky}
\bearr
|\bar M_{u\bar u\rightarrow u\bar u}|^2 &=&
s^2|D_\pi(\sqrt s,0)|^2
+t^2|D_\pi(0,\sqrt{-t})|^2
(s-4m^2)^2|D_\sigma(\sqrt s,0)|^2
+(t-4m^2)^2|D_\sigma(0,\sqrt{-t})|^2\nonumber\\
&+&\frac{1}{N_c}Re\bigg[st D_\pi^*(\sqrt s,0)D_\pi(0,\sqrt{-t})
+s(4m^2-t)D_\pi^*(\sqrt s,0)D_\sigma(0,\sqrt{-t})\nonumber\\
&+&t(4m^2-s)D_\pi(0,\sqrt{-t})D_\sigma^*(\sqrt s,0)
+(4m^2-s)(4m^2-t)D_\sigma(0,\sqrt{-t})D_\sigma^*(\sqrt s,0)\bigg].
\label{uubtouub}
\eearr
Similarly, the same for the process $u\bar d\rightarrow u\bar d$ is given as\cite{klevansky}
\bearr
|\bar M_{u\bar d\rightarrow u\bar d}|^2 &=&
4s^2|D_\pi(\sqrt s,0)|^2
+t^2|D_\pi(0,\sqrt{-t})|^2
(s-4m^2)^2|D_\sigma(\sqrt s,0)|^2
+(t-4m^2)^2|D_\sigma(0,\sqrt{-t})|^2\nonumber\\
&+&\frac{1}{N_c}Re\bigg[-2st D_\pi^*(\sqrt s,0)D_\pi(0,\sqrt{-t})
+2s(4m^2-t)D_\pi^*(\sqrt s,0)D_\sigma(0,\sqrt{-t})\bigg].
\label{udbtoudb}
\eearr
The meson propagators in the above is given by Eq.(\ref{rpaprop}) and depend on both the masses and the widths of the mesons,
depending on the medium.

The reason for doing an averaging as in Eq.(\ref{tauinv}) is due to the fact that otherwise it 
becomes numerically challenging otherwise. In certain cases, e.g. $\pi$-$\pi$ scattering within
 chiral perturbation theory, it can be numerically managed
as the corresponding scattering amplitude square $|M|^2$ occurring in Eq.(\ref{wabs}) is a polynomial function of $s,t$ variables 
\cite{goity,lang}.
On the other hand, for the processes considered here, $|M|^2$ is a nonpolynomial nontrivial function of these variables 
arising from the meson
propagators $D_M(\sqrt s,0)$ and $D_M(0,\sqrt t)$ as may be seen in Eqs. (\ref{uubtouub}) and (\ref{udbtoudb}).

\begin{figure}[t]
\vspace{-0.4cm}
\begin{center}
\begin{tabular}{c c}
\includegraphics[width=9cm,height=9cm]{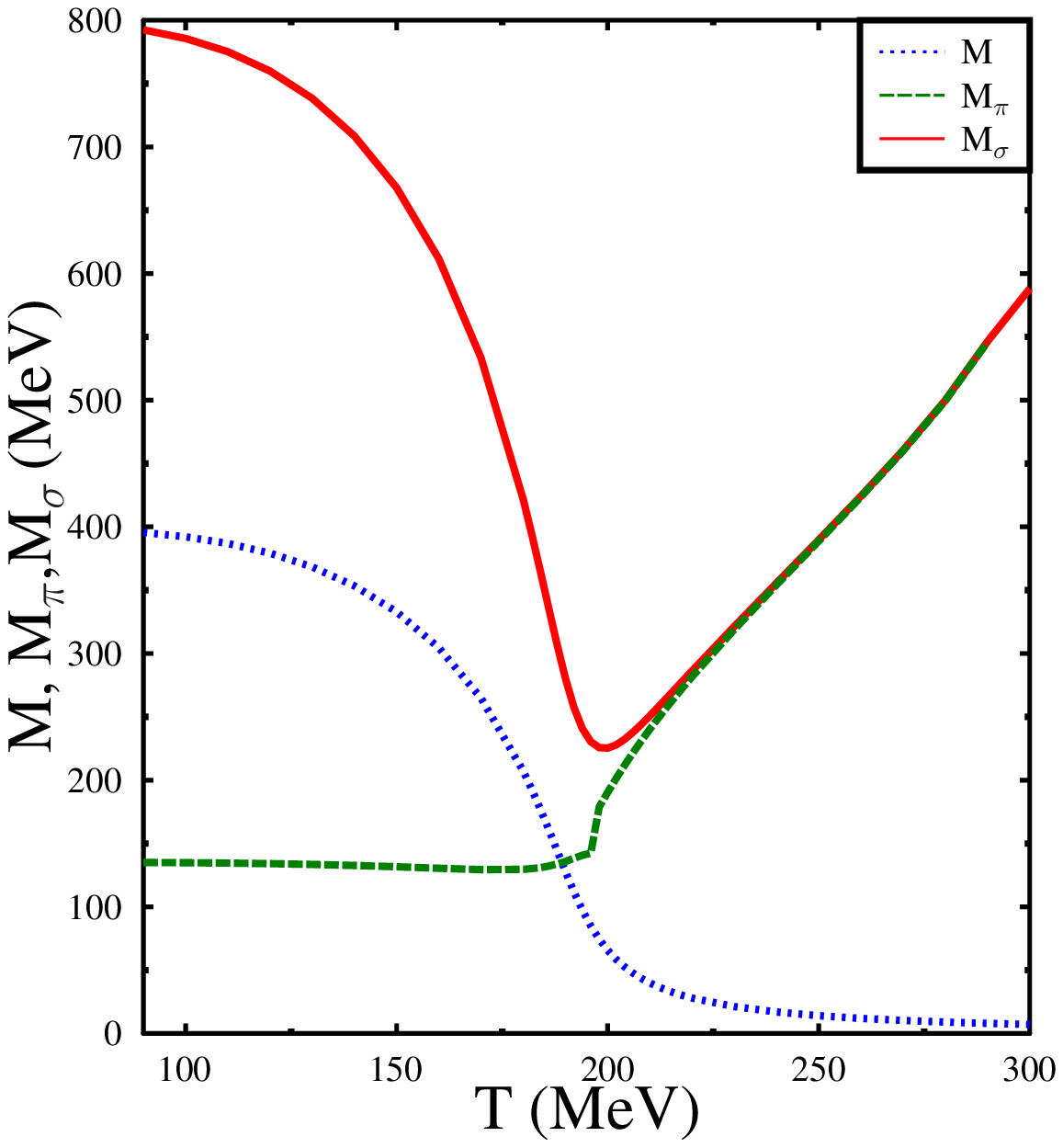}&
\includegraphics[width=9cm,height=9cm]{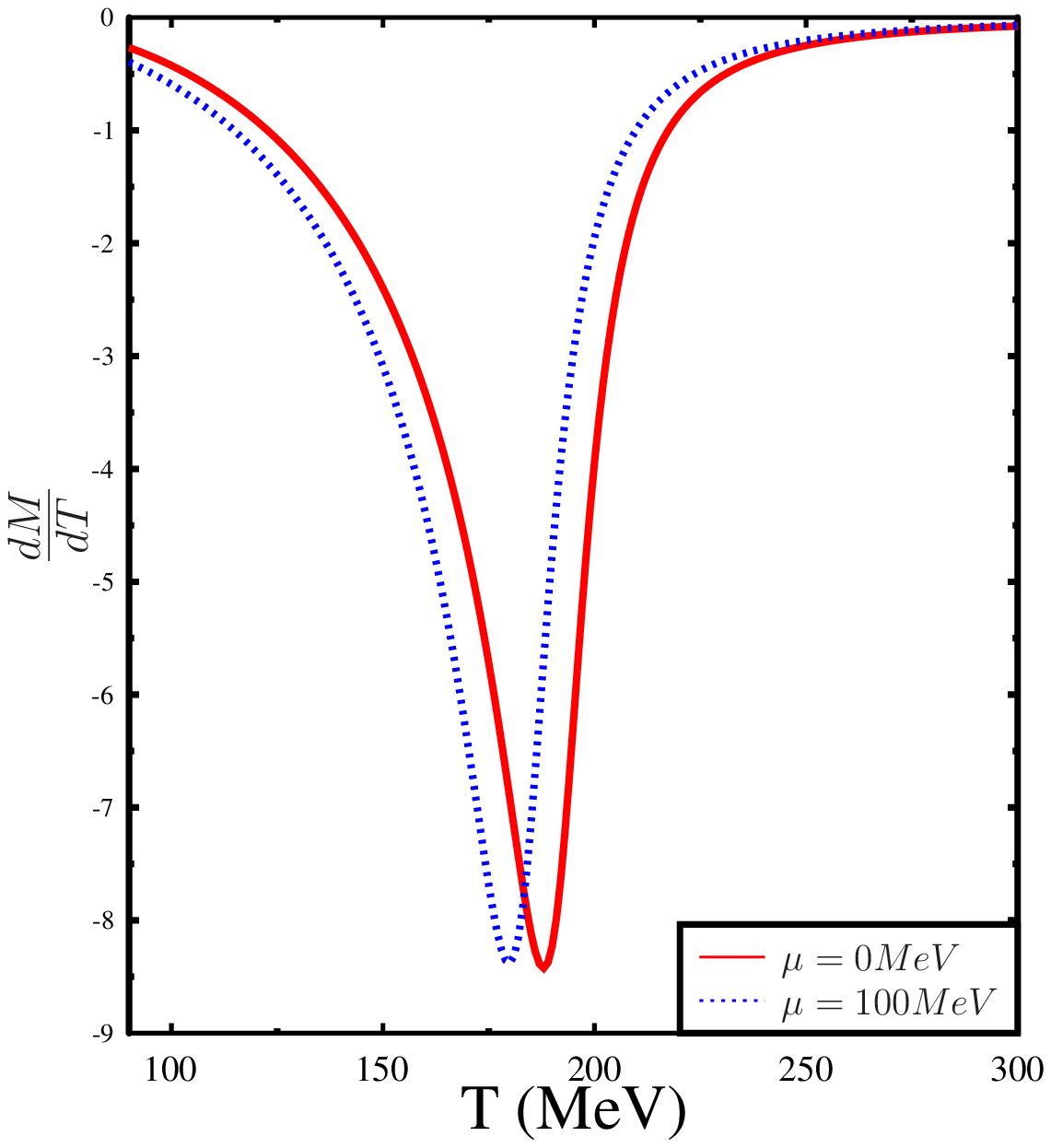}\\
Fig. 1-a & Fig. 1-b
\end{tabular}
\end{center}
\caption{ (Fig 1 a) Temperature dependence of the masses of constituent quarks ($M$), and pions ($M_\pi$) and sigma mesons ($M_\sigma$)
for $\mu=0$ and (Fig1-b)
 temperature derivative of the constituent quark mass for $\mu=0$MeV and $\mu=100$MeV .}
\label{fig1}
\end{figure}

\begin{figure}[t]
\vspace{-0.4cm}
\begin{center}
\begin{tabular}{c c}
\includegraphics[width=9cm,height=9cm]{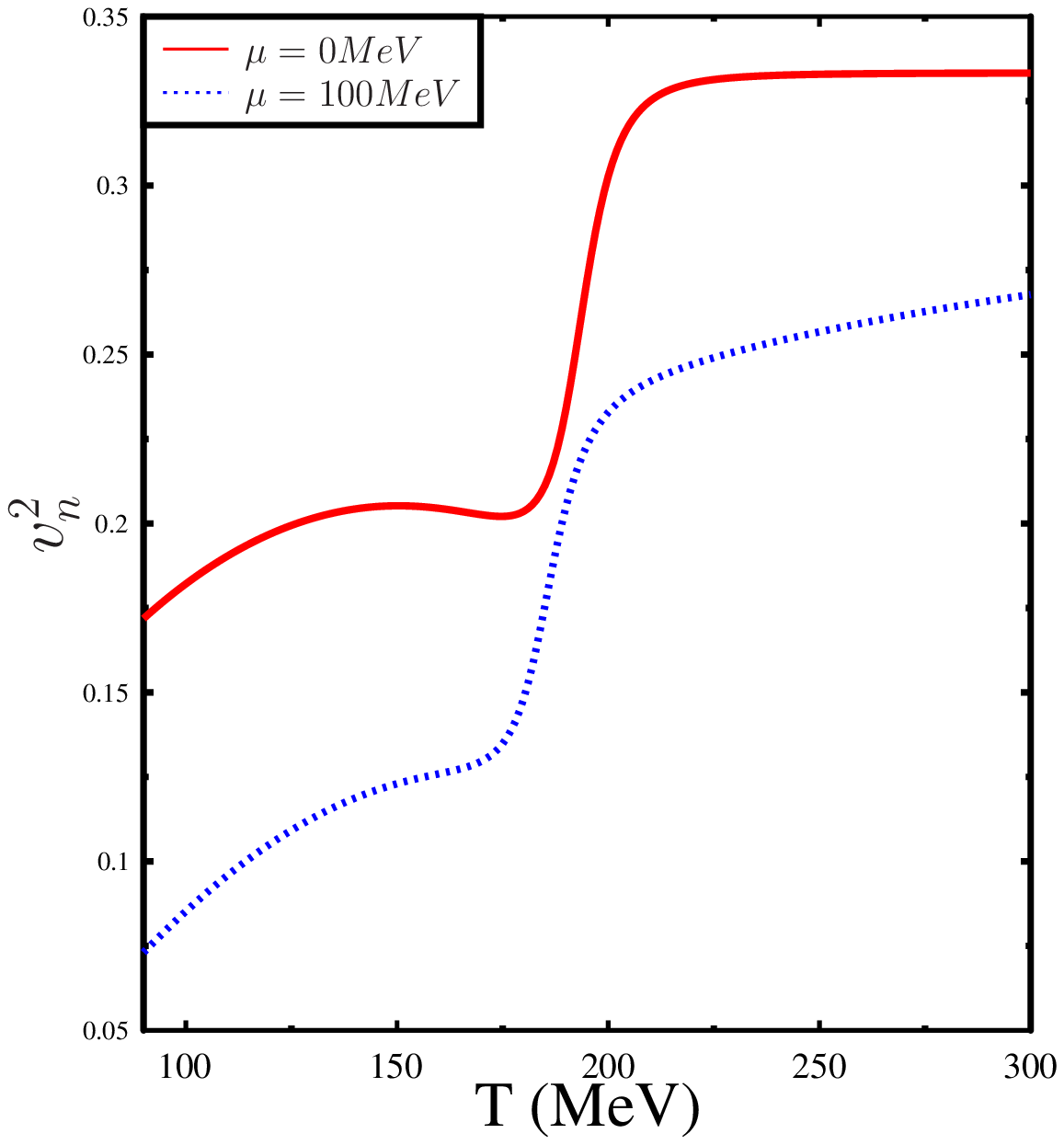}&
\includegraphics[width=9cm,height=9cm]{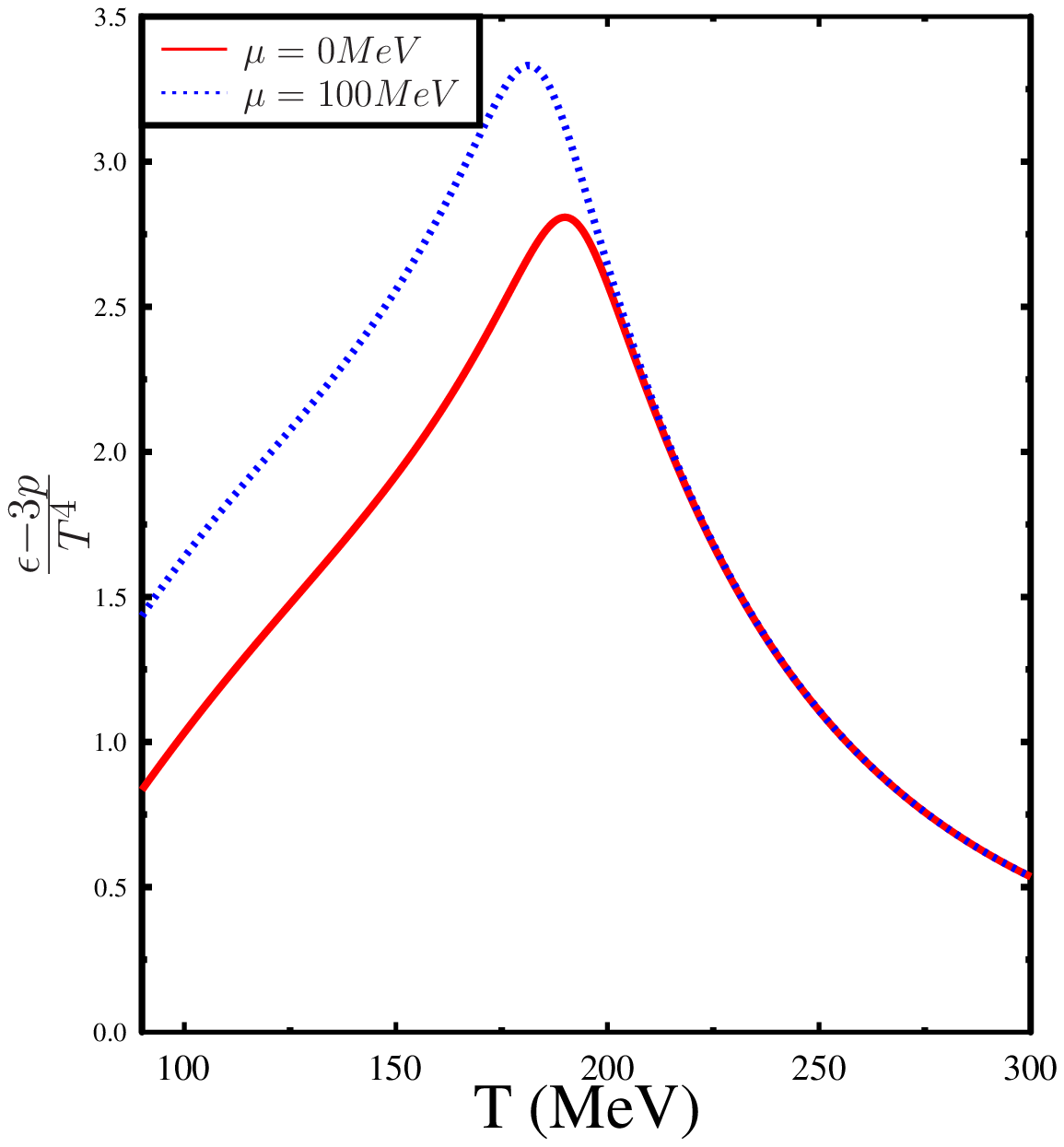}\\
Fig. 2-a & Fig. 2-b
\end{tabular}
\end{center}
\caption{  Temperature dependence of square of  the velocity of sound ($v_n^2$) (Fig2-a) 
and trace anamoly $((\epsilon-3p)/T^4)$ (Fig 2-b) for $\mu=0$ MeV and $\mu=100$ MeV}
\label{fig2}
\end{figure}
\begin{figure}[t]
\vspace{-0.4cm}
\begin{center}
\includegraphics[width=9cm,height=9cm]{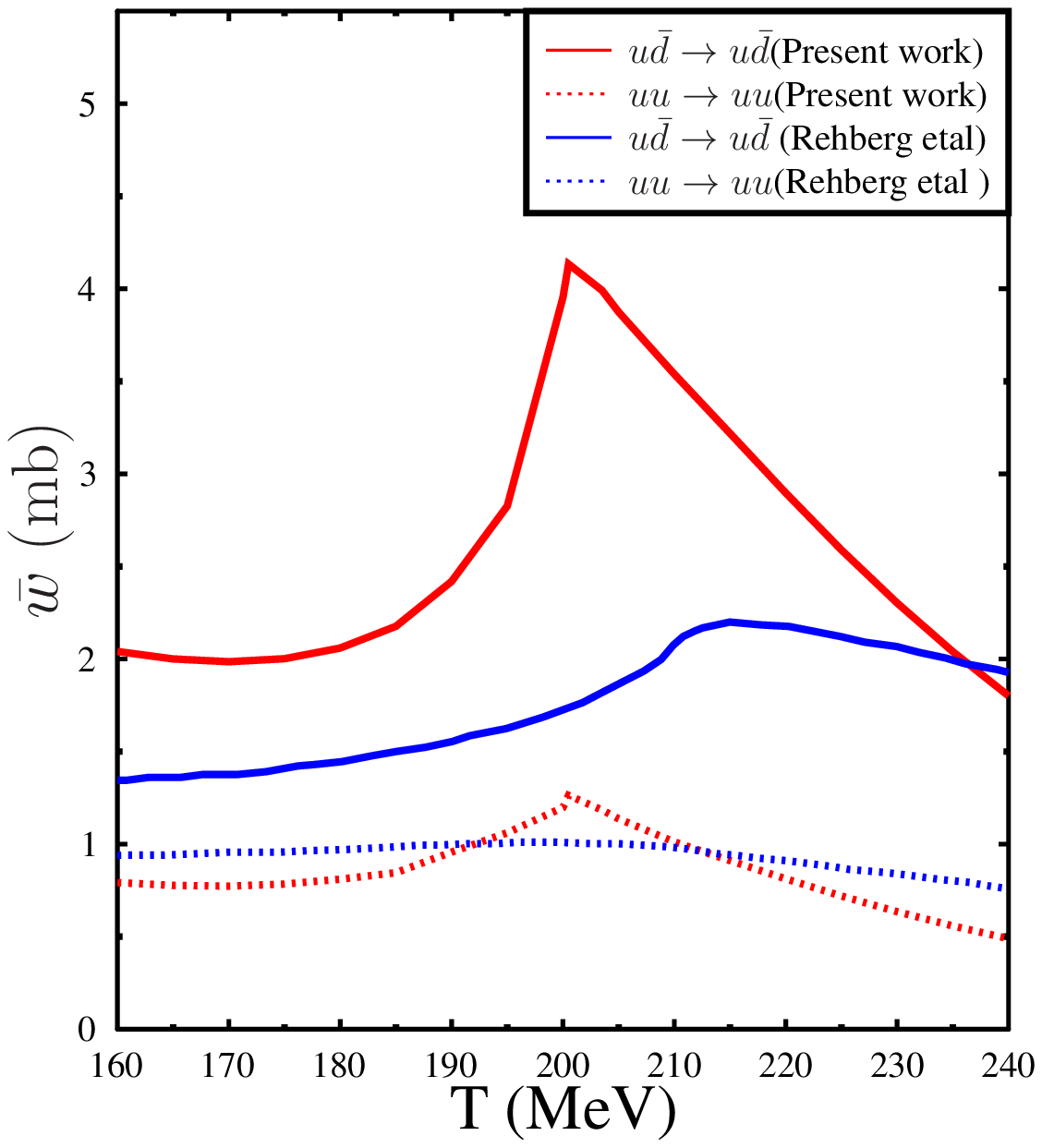}
\end{center}
\caption{Thermal averaged transition rate shown as a function of temperature. The red curves corresponds to
the present calculations while the blue curves correspond to the estimation in Ref. \cite{klevanskynpa}.
The solid and dotted lines correspond to the scattering in the quark- antiquark channels and quark quark
channels respectively.}
\label{fig3}
\end{figure}


\section{Results}

The two-flavor NJL model as given in Eq.(\ref{NJLL}), within which we shall be discussing the results, has three parameters, namely, the 
four-point coupling $G$, the three-momentum cutoff $\Lambda$ to regularize
 the integrals appearing in the mass gap equation, and, in the 
integrals involving meson masses  and the current quark mass $m$ that we take to be the same for $u$ and $d$ quarks. Within the
mean field approximation for the thermodynamic potential, and the RPA approximation for the meson
masses, these three parameters are fixed by fitting the pion mass, the pion decay constant and the quark condensate. While 
the pion mass $m_\pi=135$ MeV \cite{mpiexpt} and pion decay constant $f_\pi=92.4$ MeV\cite{fpiexpt} are known
somewhat accurately, the uncertainties in the quark condensates are rather large. Whereas, the extraction from the QCD 
sum rules turns out to be in the range 190 MeV$<-\langle\bar uu\rangle^{1/3}<$
 260 MeV ( at renormalization
scale of 1 GeV) \cite{condsum}, extraction from lattice simulation turns 
out to be $-\langle \bar u u\rangle^{1/3}\sim 231$ MeV
\cite{condlat}. Here, we have used the parameter set $m=5.6$ MeV, $\Lambda$=587.9 MeV and $G\Lambda^2=2.44$. This leads to
the vacuum value of the constituent quark mass  $M\simeq 400$ MeV, and  the condensate value is $-\langle\bar uu\rangle^{1/3}=
241$ MeV.

Let us first discuss the thermodynamics of the two-flavor NJL model as relevant for the calculation of the
transport coefficients. 

With the parameters as above, the gap equation is first solved using Eq.(\ref{gapeq})
for a given temperature and chemical potential. This is then used to solve for the masses of the pion and sigma
masses using Eqs.(\ref{pionmass}) and (\ref{sigmamass}) within the random phase approximation.
In Fig.1(a), we have plotted the constituent quark mass, and the meson masses so derived 
as a function of temperature for $\mu=0$. In the chirally broken phase, the pion mass, being the mass
of an approximate Goldstone mode is protected and varies weakly with temperature. On the other hand,
the mass of $\sigma$ ,
which is approximately twice the constituent quark mass, drops significantly near the
crossover temperature. At high temperature, being chiral partners, the masses of $\sigma$ and $\pi$ mesons become degenerate and 
increase linearly with temperature.
The constituent quark mass decreases to  small values but never vanishes.
The chiral crossover transition $T_\chi$ turns out 
to be about 188 MeV for $\mu=0$ and about 
180 MeV for $\mu=100$ MeV.  These are defined by the peak in the derivative of the
constituent mass  $(dM/dT)$, which we have shown in Fig.{(1b)}. Let us note here that the constituent mass at $T_\chi$
turns out to be about 145 MeV. On the other hand, one can have the other characteristic temperature
namely, the Mott temperature $T_M$ defined through the relation $m_\pi(T_M)=2M(T_M)$, i.e.,  the  temperature when the twice the
constituent quark mass becomes equal to that of the pion mass. 
As may be observed in Fig.1-a the 
Mott temperature for pions is about 197 MeV. This temperature is relevant in the present case where we estimate the
relaxation time using quark scattering involving meson exchange.

 Next, we  show, in Fig 2-a,  the temperature dependence of the square of the  velocity of sound $v_n^2=(dp/d\epsilon)_n$ at constant 
quark number density as defined in Eq.(\ref{vn2}).
The velocity of sound do not show any dip around the critical temperature $T_\chi$, but rises around the 
critical temperature and approaches the value of $\frac{1}{3} $ 
at high temperatures. In Fig. 2-b , we show the dependence of the 
trace anomaly $(\epsilon-3p)/T^4$. The conformal symmetry is broken maximally 
at the critical temperature and is larger for higher chemical potential.

\begin{figure}[t]
\vspace{-0.4cm}
\begin{center}
\begin{tabular}{c c}
\includegraphics[width=9cm,height=9cm]{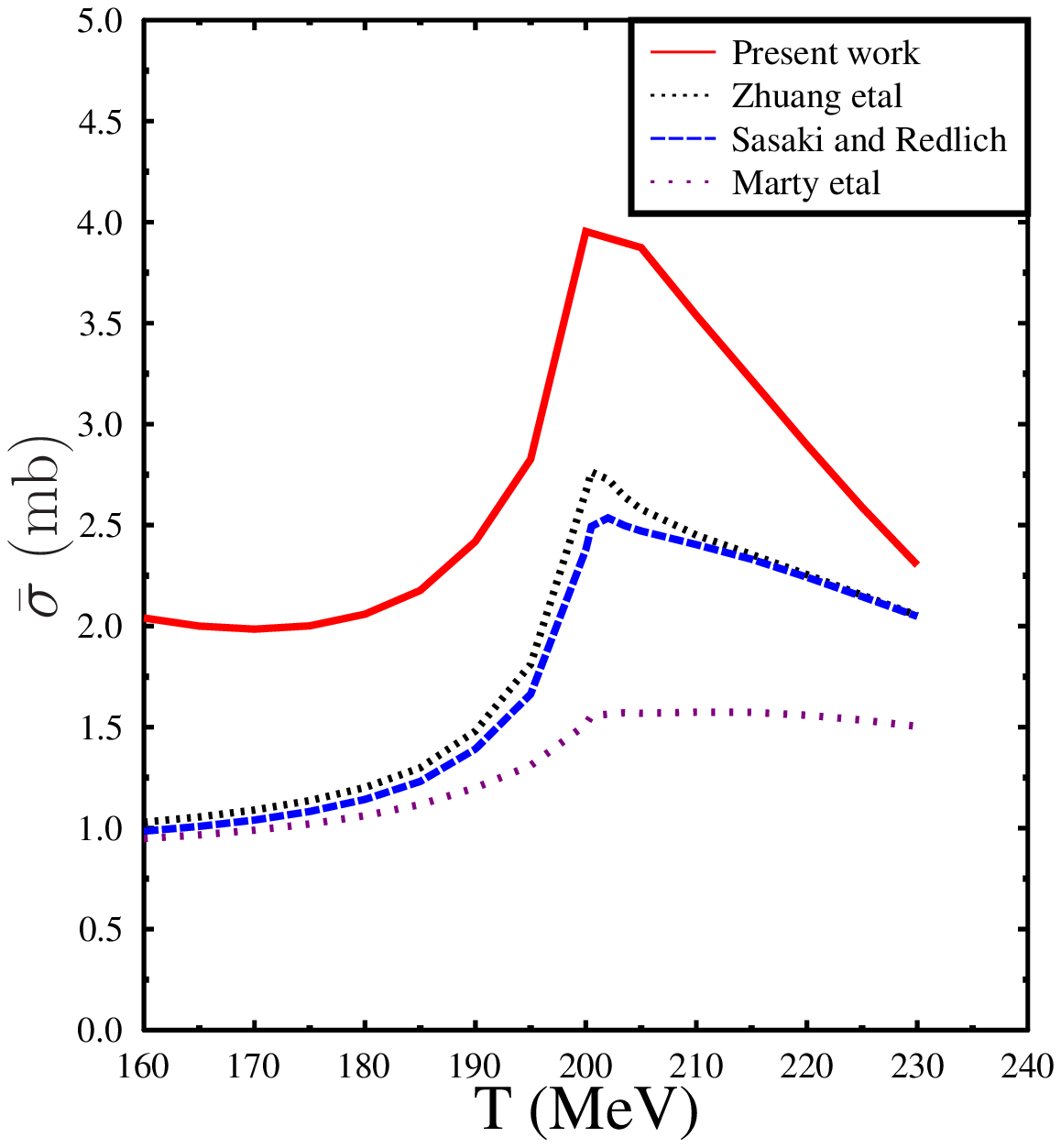}&
\includegraphics[width=9cm,height=9cm]{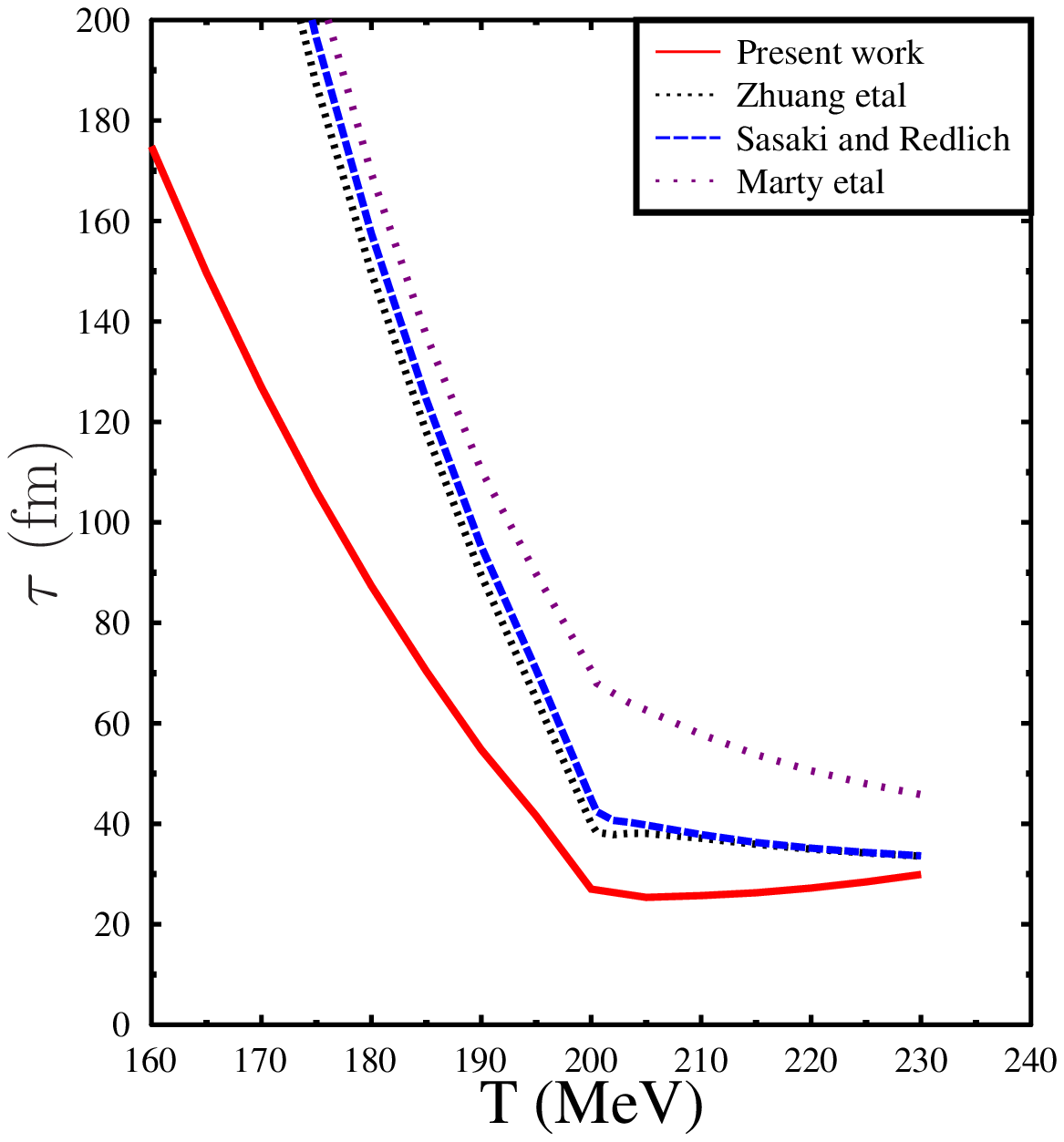}\\
Fig. 4-a & Fig. 4-b
\end{tabular}
\end{center}
\caption{ Thermal-average cross section for the process $u\bar d\rightarrow u\bar d$ 
as a function of temperature is plotted in Fig 4-a. The  present
calculation is shown by the solid line. The other results correspond to Zhuang etal \cite{klevansky},
Sasaki and Redlich \cite{sasakinjl} and Marty etal \cite{marty} . In Fig 4-b is plotted the
corresponding relaxation time $\tau_{u \bar d \rightarrow u \bar d}=\bar\sigma_{u\bar d\rightarrow u\bar d} n_{\bar d}$.}
\label{fig4}
\end{figure}

\begin{figure}[t]
\vspace{-0.4cm}
\begin{center}
\begin{tabular}{c c}
\includegraphics[width=9cm,height=9cm]{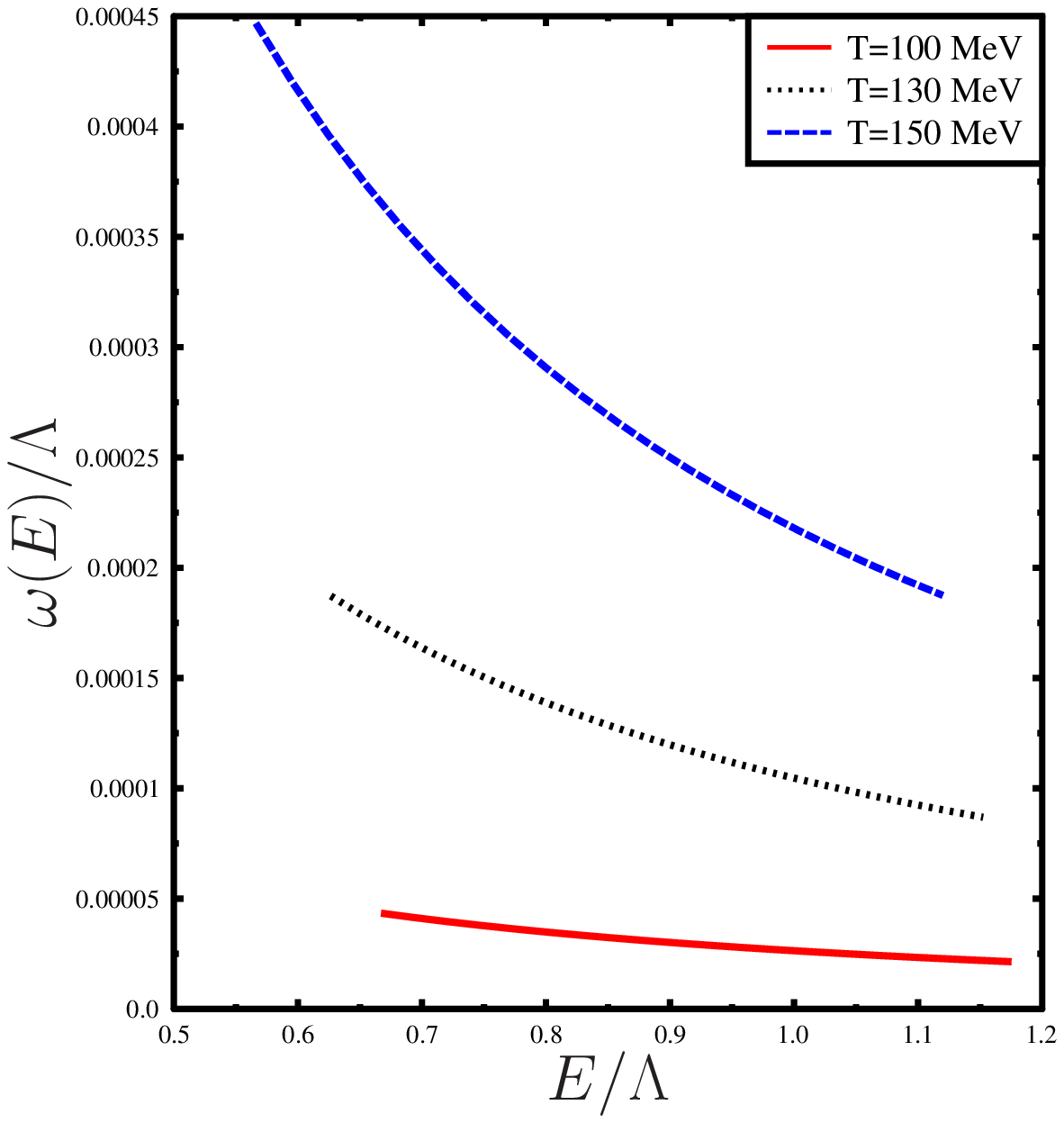}&
\includegraphics[width=9cm,height=9cm]{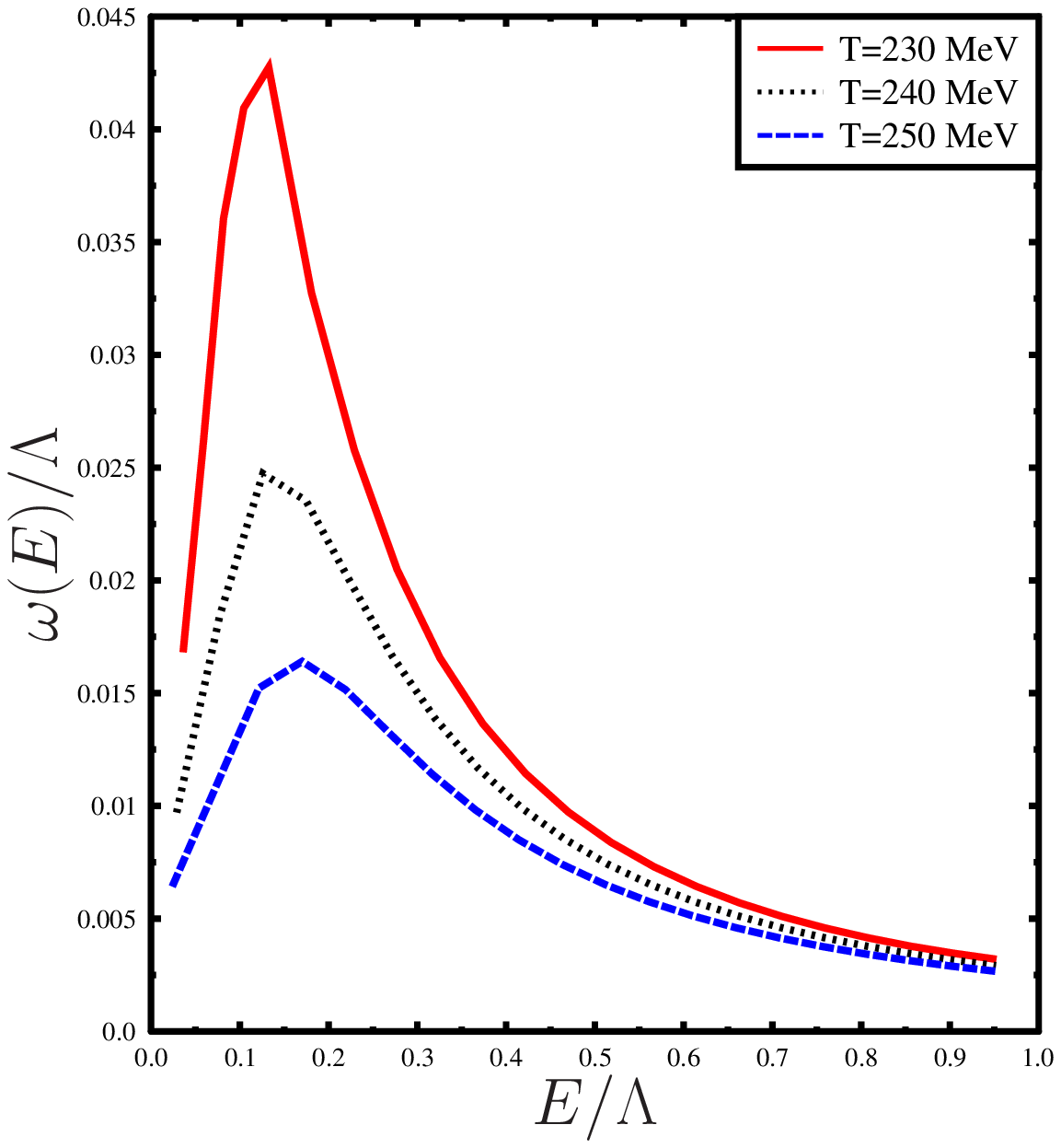}\\
Fig. 5-a & Fig. 5-b
\end{tabular}
\end{center}
\caption{ Transition frequency in units of $\Lambda$ for the process $u\bar d\rightarrow u\bar d$ 
as a function of function of energy for different temperatures. For temperature less than
the Mott temperature is plotted in Fig 5-a while for temperature higher than Mott temperatures is plotted on Fig 5-b.
}
\label{fig5}
\end{figure}
\begin{figure}[t]
\vspace{-0.4cm}
\begin{center}
\begin{tabular}{c c}
\includegraphics[width=9cm,height=9cm]{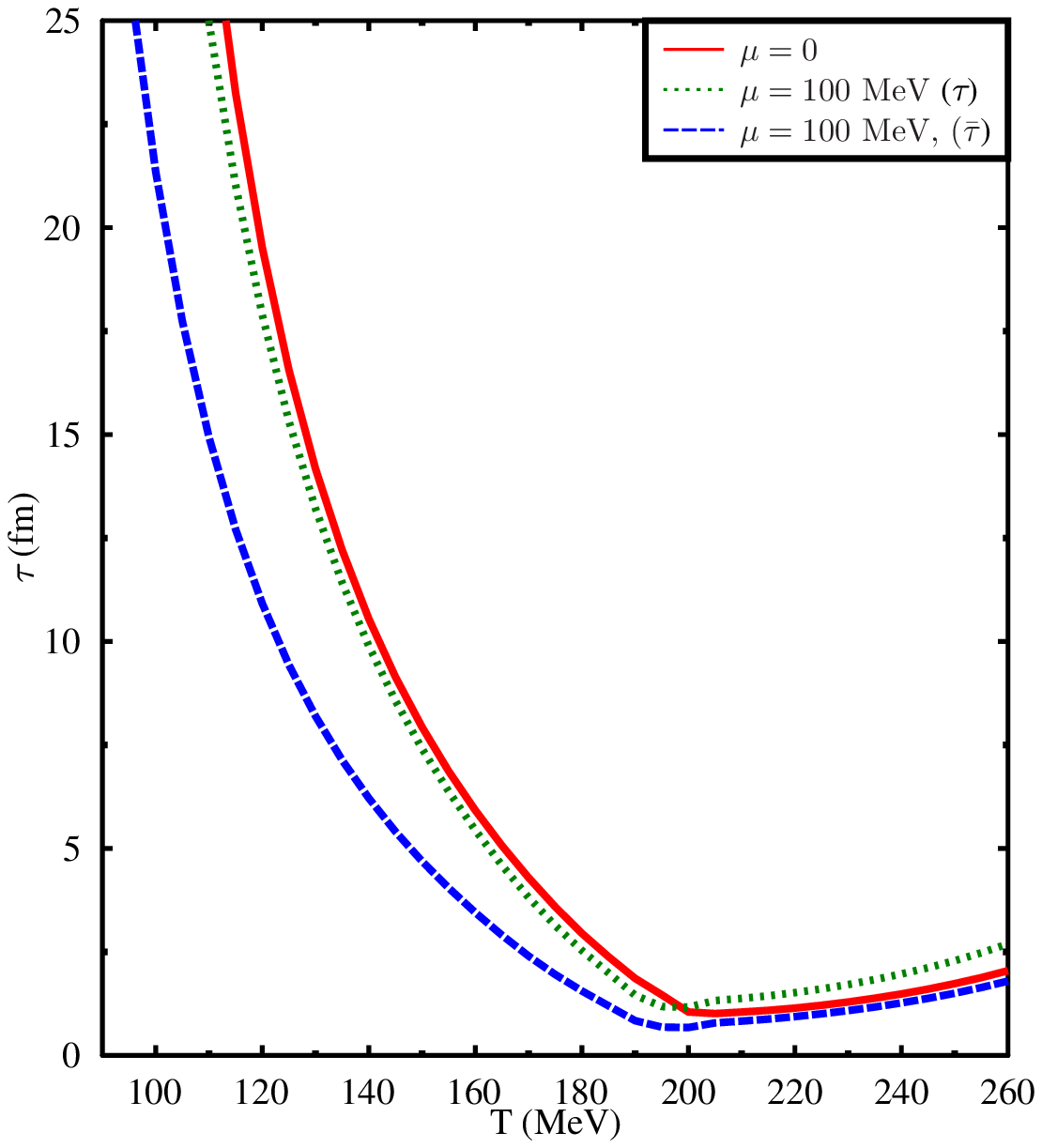}&
\includegraphics[width=9cm,height=9cm]{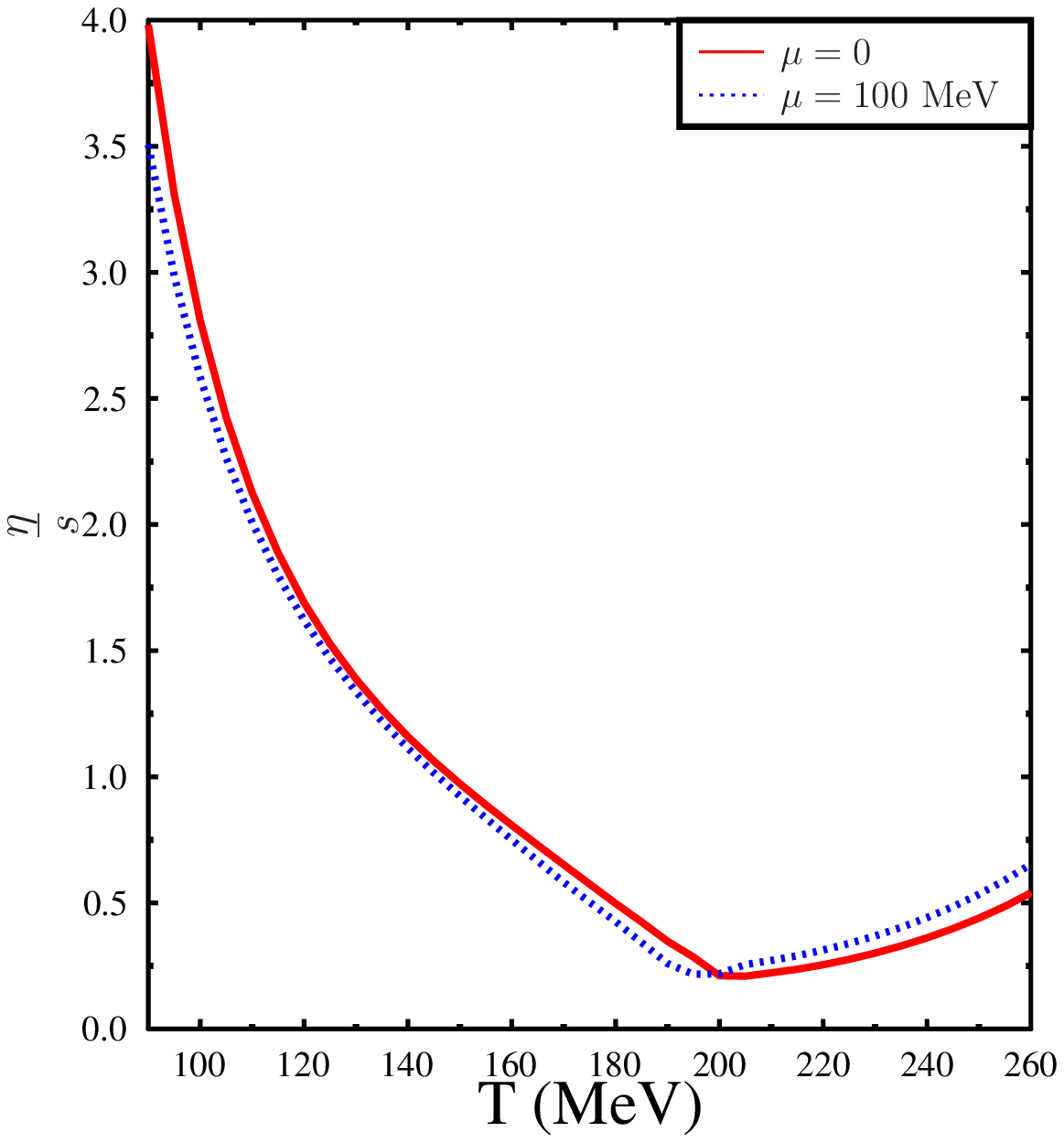}\\
Fig. 6-a & Fig. 6-b
\end{tabular}
\end{center}
\caption{Relaxation time as a function of temperature for $\mu=0$ MeV and for $\mu=100$MeV (Fig 6-a).
 In Fig (6-b), shear viscosity to entropy density ratio is shown for $\mu=0$ MeV and $\mu=100$ MeV.}
\label{fig6}
\end{figure}

We would like to mention that the behavior of the velocity of sound
shows  a different behavior as compared to 
lattice simulations \cite{tanmoy} where it shows a minimum and
 then rises to a value a
little less than the ideal gas limit of $1/3$ . The present results for the sound velocity are similar in nature to linear sigma model 
calculations of Ref.\cite{purnendu} with a lighter sigma meson 
of mass about 600 MeV. 
This behavior, as we shall observe later, gets reflected in the results
for the bulk viscosity.

We then plot the thermal-averaged transition rate $\bar W_{ab}$ of Eq.(\ref{wbarab}) for quark scatterrings for 
the quark-antiquark chaneel and quark-quark channel in Fig.\ref{fig3}. Quark-antiquark scattering turns out to be  dominant 
compared to quark-quark scattering.
After thermal averaging the transition rate shows a peak around the Mott transition temperature.
 Above the Mott temperature, the transition rates decrease with temperature. 
This one would expect as the meson propagators 
$D_M^{-1}(\sqrt{s},\zbf 0)\sim \left[(s-\left(m_M-i\Gamma/2\right)^2\right]$ and both the resonance mass and the width 
increase with temperature \cite{zhuang}.
The behavior of the transition rate is qualitatively similar to that in Ref. \cite{klevanskynpa}.
The difference could be due to the fact that in Ref. \cite{klevanskynpa}, where, three-flavors case is considered, there 
could be more channels possible
for quark scattering and also the parameters of the model are different. In the present case, the transition rate decreases 
faster as compared to Ref \cite{klevanskynpa} beyond the Mott temperature, leading to a rise of average relaxation time
 as can be expected from Eq.(\ref{tauinv}).
 
A comment regarding estimation of the mean relaxation time may be relevant here, although, ideally one would like to keep
the energy-dependent relaxation time and perform the phase space integration in the expression for the transport coefficients. 
In the present calculations, this is carried out by calculating a  mean interaction frequency of the energy dependent
interaction frequency related to the standard quantum field theoretic transition rate as in Eqs.(\ref{tauinv}), (\ref{omgea})
and (\ref{wabs}). On the other hand, in Ref.s 
\cite{klevansky,sasakinjl,marty} this  averaging is performed by first obtaining an energy averaged cross section e.g. for the
process $u,\bar d\rightarrow u,\bar d$ 
\be
\bar\sigma_{u\bar d\rightarrow u\bar d}(T,\mu)=\int ds \sigma_{u\bar d\rightarrow u\bar d}(T,\mu) P(s,T,\mu)
\label{prosig}
\ee
where, $P(s,T,\mu)$ is the probability of yielding a quark-antiquark pair with energy $\sqrt s$ and is normalized as
\be
\int ds P(s,T,\mu)=1.
\label{pronorm}
\ee
  However, in the three references cited above, the expressions for $P(s,T,\mu)$ are different. In the earliest one \cite{klevansky},
Zhuang {\it et al} take
\be
P_{zhuang}(s,T,\mu)=C\sqrt{s(s-4m^2)}f(\sqrt s/2-\mu)f(\sqrt s/2+\mu)
\label{zhuang}
\ee
 while Refs.\cite{sasakinjl} and \cite{marty} consider, respectively, the probability function as
\be
P_{sas}(s,T,\mu)=C(s-4m^2) f(\sqrt s/2-\mu)f(\sqrt s/2+\mu),
\label{sasaki}
\ee
\be
P_{marty}(s,T,\mu)=C (s-4m^2)\sqrt{s} f(\sqrt s/2-\mu)f(\sqrt s/2+\mu).
\label{marty}
\ee
In Eq.s (\ref{zhuang},\ref{sasaki},\ref{marty}), the constant $C$ is fixed from the normalization condition
of Eq.(\ref{pronorm}).
The contribution of this averaged cross section to the relaxation time  is given as
\be
\tau_{u\bar d\rightarrow u \bar d}^{-1}=n_{\bar d}\bar\sigma_{u\bar d\rightarrow u\bar d}
\ee

The resulting energy-averaged cross section as well as the corresponding relaxation time for the different assumptions
for the probability function as compared to the present averaging procedure is shown in Fig.\ref{fig4}. The general behavior of the cross section of having a peak around Mott transition is seen all the figures. However, the sharp fall of the cross section beyond the Mott transition is seen with the present
averaging while the same for  Ref.\cite{marty} is rather slow. This gets reflected in the behavior of the relaxation 
time for this process in Fig 4 b. In particular the relaxation time corresponding to Ref\cite{marty} shows a monotonic decrease beyond the Mott temperature.
In this context, it is also relevant to  analyze whether the function $\tau(E)$ is reasonably smooth so that the energy
averaging is a reasonable approximation. To verify this we also examine
 the energy-dependent interaction frequency $\omega(E)$ of Eq.(\ref{omgea}),
which explicitly simplifes to
\be
\omega_a(E_a)=\frac{1}{8\pi^2E_a}\int_{m_b}^\infty dE_b f(E_b)\int_{-1}^1 d x W_{ab} (s)
\label{omegaea}
\ee
where, --$ x=\cos\theta$.  To evaluate the above integral, we note that
\be
s-(p_a+p_b)^2=2 E_aE_b\left(\frac{s}{2E_aE_b}-y(x)\right)
\ee
with 
\be
y(x)=1+\frac{m_a^2+m_b^2}{2E_aE_b}-\frac{|p_a||p_b|}{E_aE_b}x
\ee

Inserting the identity involving the delta function
\be
1=\int ds \frac{1}{2E_aE_b}\delta\left(\frac{s}{2E_aE_b}-y\right)
\ee
in the integral Eq.(\ref{omegaea}), we have
\be
\omega(E_a)=\frac{1}{8\pi^2 E_a}
\int dE_b p_b E_b\int dx W_{ab}(s)|_{s=2E_aE_b(1+\frac{m_a^2+m_b^2}{2E_aE_b}-\frac{|p_a||p_b|}{E_aE_b}x)}.
\ee

Using Eq.(69) for the transition rate $W_{ab}(s)$, we have calculated the energy-dependent transition frequency $\omega(E_a)$ for
a typical scattering process ($u\bar d\rightarrow u\bar d$) and have plotted it in Fig 5 a and Fig 5b for temperatures below and 
above the Mott temperatures respectively.
As may be seen the energy dependence of the transition frequency is indeed a smooth function of energy. Below the critical temperature,
with an increase in temperature, the interaction frequency increases essentially due to the fact that 
the sigma meson mass decreases with temperature.
This behavior of $\omega(E)$ with energy arising from quark scattering is different from  interacting pion
 gas which diverges at large energy. Indeed, the energy-dependent
relaxation time , [inverse of $\omega(E)$]  arising from pion scattering shows a peak structure at lower energy as shown in Ref \cite{goity}.
Since pions are not the elementary degrees of freedom in the NJL model, there is no elementary pionic interaction within the model.
For  different possible behavior
of $\omega(E)$ to have finite shear viscosity  we refer to Ref.\cite{langweise}. 
On the other hand, for temperatures
above the Mott temperature,  as may be seen in Fig 5 b, the transition frequency weights the lower energy strongly. When
the temperature increases,
the peak value of $\omega(E)$ is reduced somewhat and the decay from the peak value  is slower. 
The decrease of transition frequency with temperature
beyond the Mott temperature is associated with the fact that the meson masses increase with temperature.
The rise of the transition frequency at low energies  near the Mott temperature also demonstrates the limitation of
the approximation of taking an average transition frequency. However, 
for the estimation of the transport coefficients, this low-energy peak does not make the estimation worse as it is
multiplied by a function that
itself is suppressed at low  momenta as may be seen from expression for e.g. shear viscosity in Eq.(\ref{defeta}).

\begin{figure}[t]
\vspace{-0.4cm}
\begin{center}
\begin{tabular}{c c}
\includegraphics[width=9cm,height=9cm]{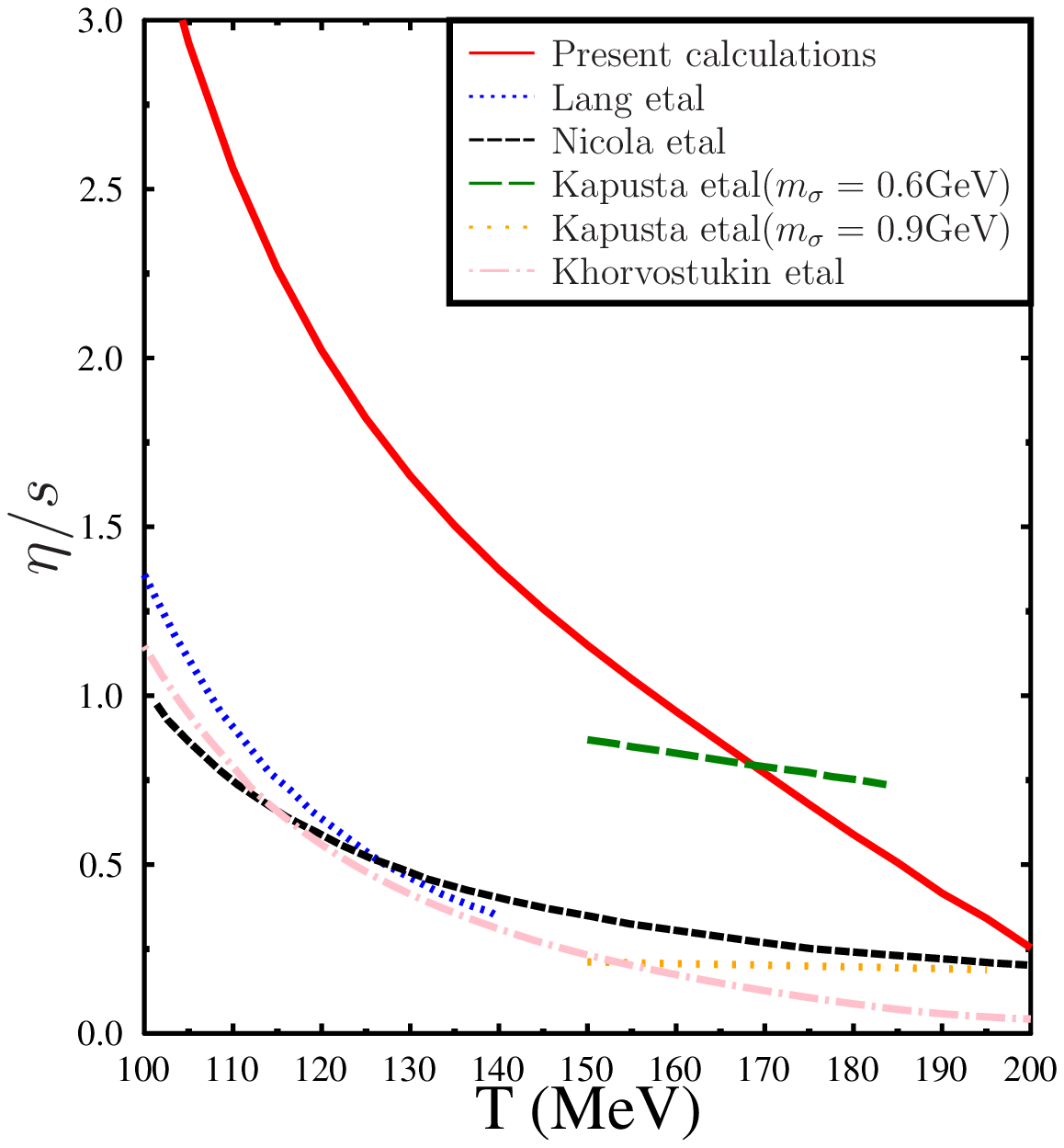}&
\includegraphics[width=9cm,height=9cm]{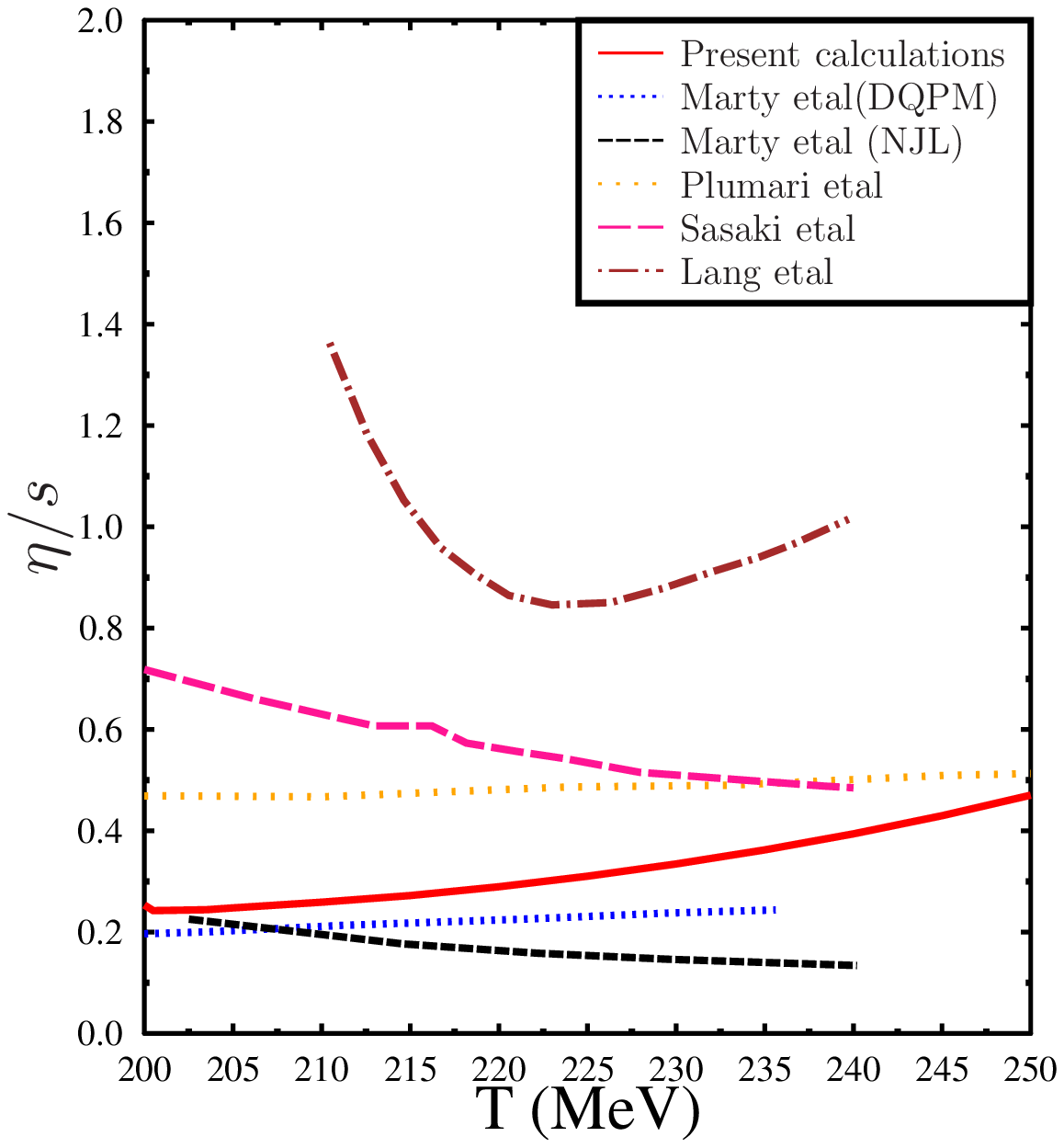}\\
Fig. 7-a & Fig. 7-b
\end{tabular}
\end{center}
\caption{Shear viscosity to entropy ratio for $\mu=0$ for low temperatures (Fig 7-a). The present calculations 
is shown by the solid lines. The other results correspond to Lang etal \cite{lang} of interacting pion gas,
Fernandez-Fraile and Nicola\cite{nicola}. The two curves by Kapusta etal from Ref.\cite{purnendu} correspond to
different masses for the sigma mesons. The green dashed curve is for $m_\sigma=600$ MeV while the orange dotted curve is with 
$m_\sigma=0.9$ GeV. The pink dot dashed curve is for the SHMC model by Khvorostukhin etal of Ref\cite{voskresensky}.  
In Fig. 7-b  is shown the ratio for higher temperatures. Present calculations is shown by solid red line, 
the two curves of Marty etal \cite{marty}
correspond to dynamical quasi particle model (DQPM) and the 3 flavor NJL model, the orange dotted curve by Plumari etal is from Ref.
\cite{greco} , the pink dashed curve by Sasaki etal is for two flavor NJL model of Ref. \cite{sasakinjl} while the brown dot dashed curve b Lang etal is 
from Ref.\cite{langweise}.
 }
\label{fig7}
\end{figure}

\begin{figure}[t]
\vspace{-0.4cm}
\begin{center}
\begin{tabular}{c c}
\includegraphics[width=9cm,height=9cm]{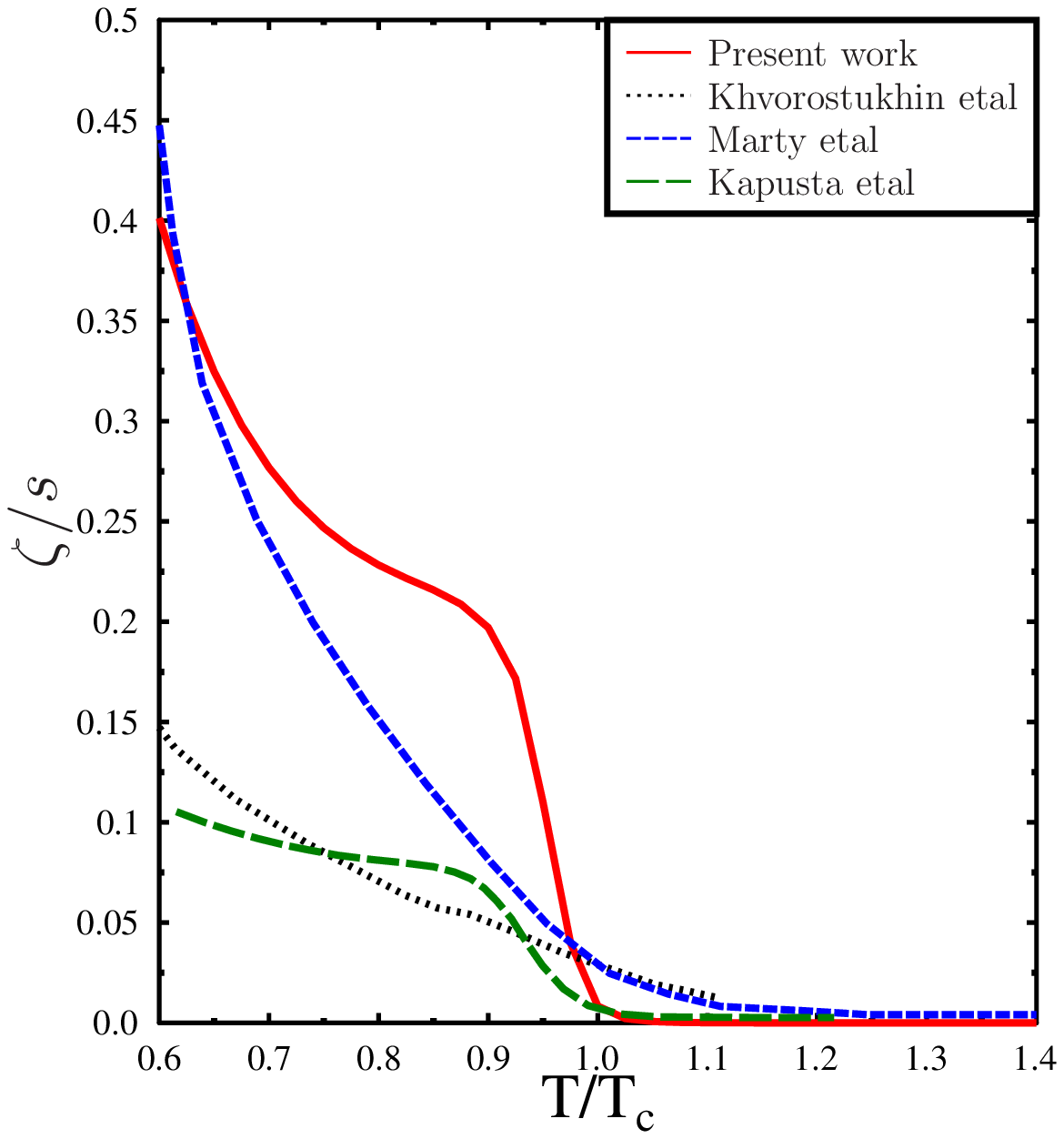}&
\includegraphics[width=9cm,height=9cm]{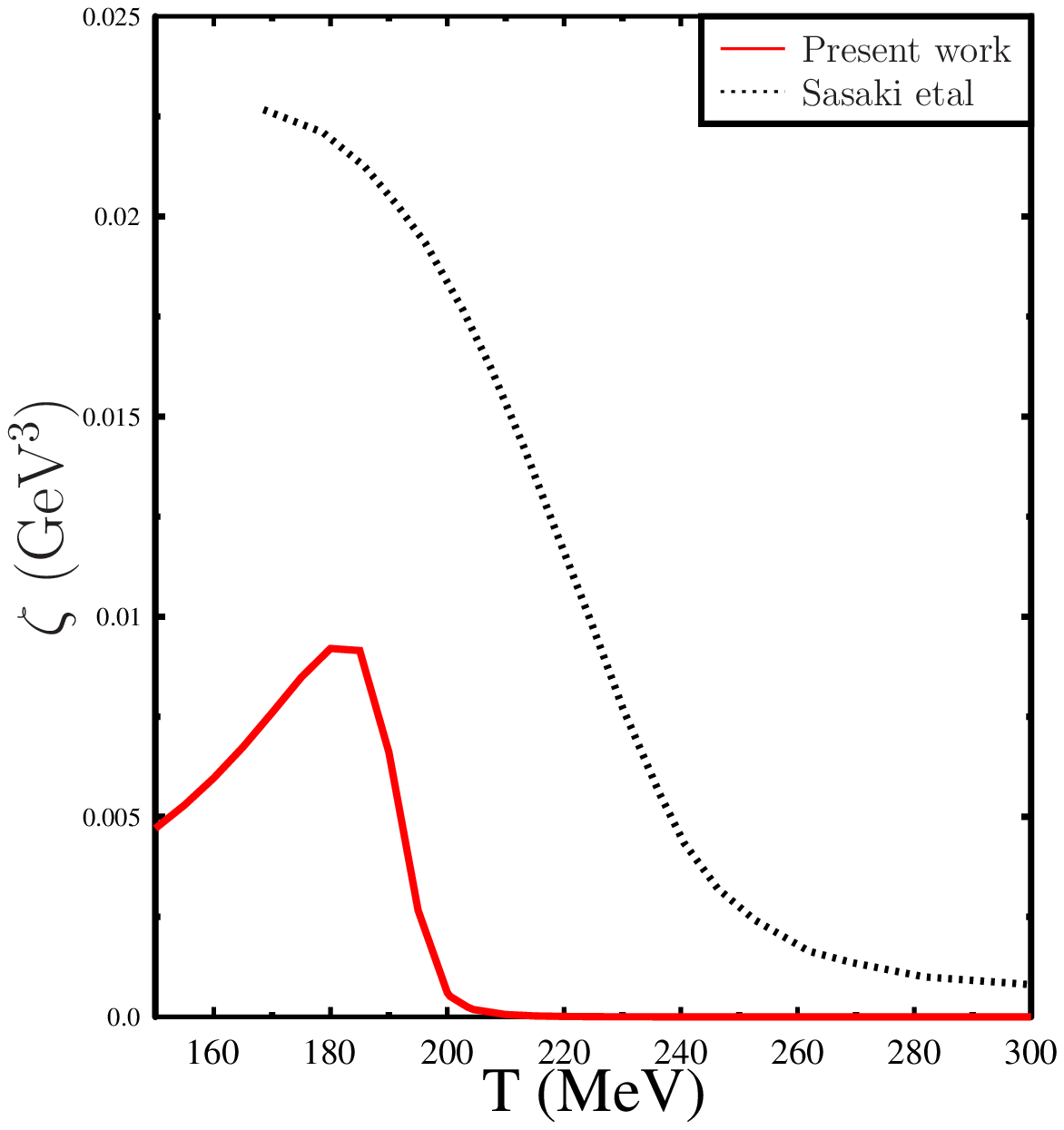}\\
Fig. 8-a & Fig. 8-b
\end{tabular}
\end{center}
\caption{ Bulk viscosity to entropy ratio as a function of temperature in units of $T_c$ for zero baryon density (8-a). 
Also shown results from different   models, the SHMC model of Khvosthukin etal\cite{voskresensky}, Kapusta etal \cite{purnendu},
the three flavor NJL results of Marty etal\cite{marty}. Bulk viscosity in units of GeV$^3$ as a function of temperature
is shown in Fig. 8-b. Solid red curve corresponds to the present calculations while
dotted curve correspond to the results by Sasaki etal \cite{sasakinjl}.
 }
\label{fig8}
\end{figure}

Next, with this limitation close to the Mott temperature, we 
 discuss the estimation of averaged relaxation time from all the scatterings  in the present approach 
as a function of temperature. Let us recall that 
this quantity is inversely related to the transition rate $\bar W_{ab}n_b$  as in Eq.(\ref{tauinv}) where $W_{ab}$is the transition rate of
all processes with species $a$,$b$ in the the initial states and is related to the corresponding scattering cross section as in Eq.(\ref{wsig}).
In general, the dominant contribution here comes from quark-antiquark scattering from the $s$ channel through propagation of 
the resonance states, the pions and the sigma. The mass of the sigma meson decreases with an increase in temperature, becoming 
a minimum at the Mott transition temperature $T_M$ and leading
to an enhancement of the cross section. This, in turn, leads to a minimum in the  relaxation time. Beyond the transition temperature
the resonance masses increase with temperature linearly leading to a smaller cross section and hence an increase
in the relaxation time beyond the Mott temperature. 
This generic feature is observed in Fig.6-a.

Let us note that $\tau^a$ depends both on the transition rate and the density of the particles of the initial state other than the species 
$a$. It turns out that the transition rate is dominant for the process $u\bar d\rightarrow u\bar d$. At finite chemical potential, 
for temperatures greater than the transition temperature, quark density is larger compared to the antiquarks. As there are fewer anti quarks 
to scatter off , the cross section for quark-antiquark scattering decreases leading to $\tau(\mu)> \tau(\mu=0)$. On the other hand,
for anti-quarks, there are more quarks to scatter off at nonzero $\mu$ as compared to at $\mu=0$. This leads to a lower 
 value for relaxation time for the antiquark at finite $\mu$ as compared to $\mu=0$ case. On the other hand, for temperatures below the 
critical temperature, while the quark-antiquark transition rate is dominant, the density of antiquark is suppressed very 
much by the constituent quark mass  for $\mu\neq 0$. The quark number density however is enhanced and the  
contribution from quark quark scattering becomes
more important resulting in a smaller value for the relaxation time at finite $\mu$ compared to the $\mu=0$ case. 

 At this stage, perhaps it may be relevant to discuss the validity of the Boltzmann kinetic approach that has been used to estimate the 
transport coefficients within the relaxation time approximation. The same can be a reliable approximation
provided the mean free path $\lambda$ is  larger than the range $d$. 
One can define the average mean free path as $$\lambda_f= v_f\tau_f$$ for a given flavor $f$. . Here the mean velocity $ v_f$ is given by
\be
v_f=\frac{2N_c}{(2\pi)^3 n_f}\int\frac{d\zbf p}{E_{\zbf p}}|\zbf p|f(E_{\zbf p}).
\ee

It turns out that at $T_{Mott}$, $\lambda_f= 1.2 fm$. At the same temperature, the mass of the pion or sigma meson turns out to be 
about 200 MeV with the corresponding Compton wavelength to be about a Fermi so that the value of the ratio $\lambda/d$ is about 1.2.
This ratio is minimal at the Mott temperature and increases rapidly both below and above the Mott temperature. Thus, within the NJL model,
it is not too much unreliable to use the Boltzmann equation within the relaxation time approximation except at the Mott transition temperature.
Keeping this in mind, we next proceed to estimate the transport coefficients.

In Fig(6-b) we have plotted the shear viscosity to entropy ratio ($\frac{\eta}{s}$) as a function of temperature
for $\mu=0$ MeV and $\mu=100$ MeV . As expected from the $\tau$ behavior with temperature, $\eta/s$ has a minimum
with $\eta/s|_{min}\sim 0.24$ at the critical temperature beyond which it increases slowly. This behavior of having a minimum around the Mott transition 
due to the suppression of scattering cross sections at higher temperatures is in contrast to results of Ref.\cite{marty} where it
shows a monotonic decrease with the value of the ratio going below the KSS bound. 
At finite $\mu$, the ratio $\eta/s$ is larger as compared to vanishing $\mu$.
This is due to two reasons. Firstly, $\tau$ at finite $\mu$ is larger and, further,
the quark density is also  larger
as compared to the antiquarks at finite density.

We also compare our results for $\eta/s$ with different existing results for  low temperatures below the
Mott transition based on different hadronic models for vanishing baryon chemical potential in Fig(7-a).
These include results within a interacting pion gas model
by Lang etal\cite{lang},  models based on chiral perturbation theory by Fernandez-Fraile and Nicola \cite{nicola},
linear sigma model within relaxation time approximation \cite{purnendu} and a hadronic model based on
scaling of hadronic masses and couplings (SHMC) by Khvorostukhin etal \cite{voskresensky}. In general, the behavior
of $\eta/s$ seems to be in conformity with these models with the ratio monotonically decreasing with temperature
in the range of temperatures considered. In all these models the dominant contributions
arise from the pions. The decreasing behavior of the ratio in the present NJL model, on the other hand, arises from the scattering of the 
constituent quarks exchanging mesons whose mass decreases with temperature leading to a decreasing behavior of the relaxation time.
On the other hand, the present estimation of  $\eta/s$ overestimates the results of  all the models shown in the figure.
This could be due to the fact that within the NJL model, the degrees of freedom at these temperatures is a system of 
constituent quarks rather than pions which should be the dominant physical degrees of freedom. This is also reflected in the
behavior of the energy-dependent interaction frequency, which is very different compared to 
the same due to pion scattering\cite{langweise,goity}.  Apart from this,  a somewhat larger value for the ratio within the 
present model could be due to the small values for the entropy density
corresponding to the heavy constituent quarks as compared to the pions.

In Fig. (7-b) we compare our results for $\eta/s$ with those of other models for higher temperatures. In contrast to the results in
Ref.s \cite{sasakinjl} and \cite{marty}, the ratio does not show a monotonically decreasing behavior. The minimum value value of
$\eta/s=0.24$ much above the KSS bound around $T_{Mott}$  and beyond which it increase monotonically. Such  a rising behavior with temperature
is also seen in two-flavor NJL model using Kubo formalism in Ref\cite{langweise}.

In Fig. (8-a) we  have plotted the specific bulk viscosity normalized to entropy density as a function of temperature.
We have also shown here the results of earlier calculations, based on the  linear sigma model \cite{purnendu}, NJL model \cite{marty}
and SHMC model \cite{ voskresensky}. 
The ratio of bulk viscosity to entropy density
increases rapidly near the critical temperature as  temperature decrease from high temperature beyond the critical temperature 
to temperatures below it. However, it is not a maximum at the critical  temperature. After the rapid rise near the  critical temperature
it increases slowly. As may be observed, in all these calculations the ratio $\zeta/s$ decreases monotonically with temperature. 
 We might mention here that, such a behavior of decreasing bulk viscosity to entropy ratio was also observed in estimations based on
PHSD transport codes \cite{phsdbrat} as well as in the linear sigma model in the large N limit \cite{dobado}. 
 This is in contrast with results in  Refs. \cite{karschkharzeev,kharzeevtuchin,latticemeyer} where $\zeta/s$
shows a peak near the critical temperature. On the other hand, we have also plotted the bulk viscosity in units of $GeV^3$ in Fig. (8-b) 
where it shows a maximum around
the Mott temperature. Such a feature of a maximum was also seen for the NJL model for two flavors 
in Ref. \cite{sasakinjl}. Such a peak in $\zeta$ was also observed in Ref.
\cite{nicolaprl} within a chiral perturbation theory framework with a maximum value of about $\zeta\sim 0.008$ GeV$^3$ as compared to
$\zeta\sim 0.01 $GeV$^3$ in the NJL model here.  However, the ratio $\zeta/s$ does not show such a
peak, probably because of the fact that the entropy of the system with massive constituent quarks
 becomes rather small to
mask the peak structure in $\zeta$. 

Beyond the Mott transition temperature the ratio $\zeta/s$ vanishes. Let us note that 
one can rewrite the expression for the bulk viscosity as
\bearr
\zeta&=&
\frac{1}{9T}\sum_a\int d\Gamma^a\frac{\tau^a}{E_a^2}f_a^0(1-f_a^0)\nonumber\\
&\times&\left[\zbf p^2(1-3 v_n^2)-3v_n^2\left(M^2-
TM\frac{dM}{dT}-\mu M\frac{dM}{dT}\right)+
3 \left(\frac{\partial P}{\partial n}\right)_\epsilon\left(M\frac{ dM}{\partial\mu}-E_at^a\right)\right]^2\nonumber\\
\label{zeta3}
\eearr

\begin{figure}[t]
\vspace{-0.4cm}
\begin{center}
\begin{tabular}{c c}
\includegraphics[width=9cm,height=9cm]{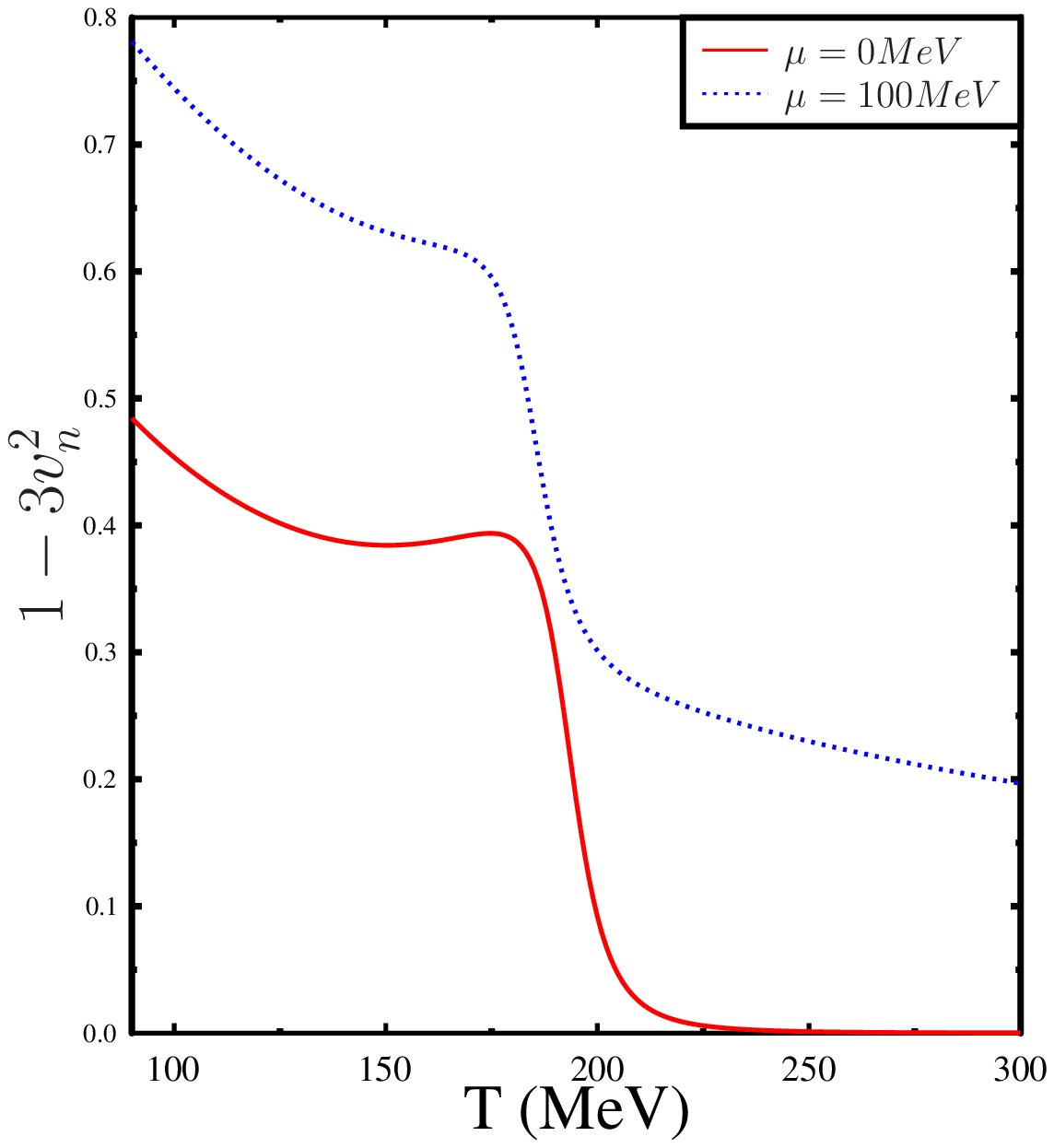}&
\includegraphics[width=9cm,height=9cm]{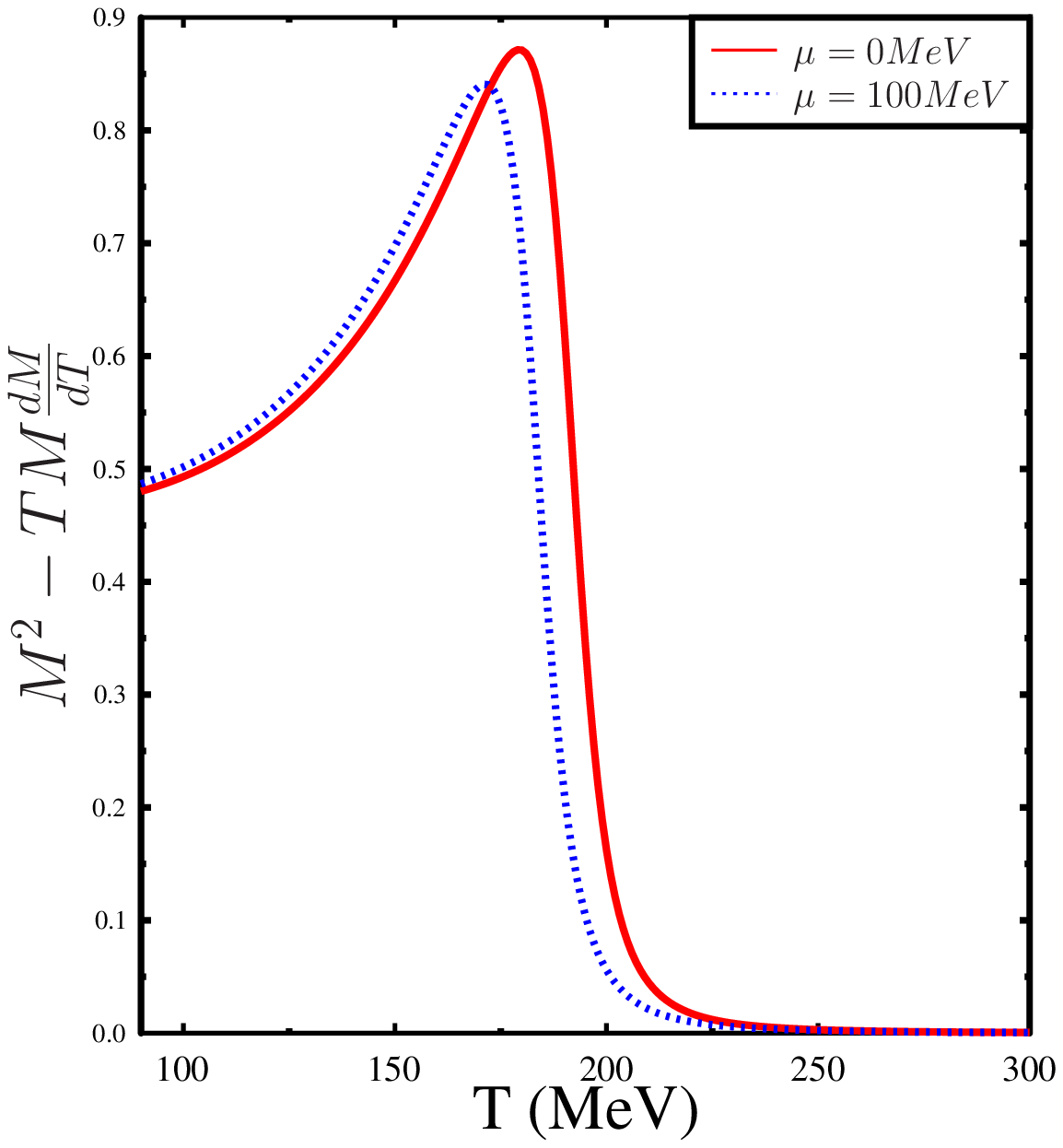}\\
Fig. 9-a & Fig. 9-b
\end{tabular}
\end{center}
\caption{the violation of conformality measure $C_1=1-3v_n^2$ (Fig 9a) and $C_2=M^2-TM\frac{dM}{dT}$ (Fig 9b) as a function of temperature
 for $\mu=0$ MeV and for $\mu=100$ MeV}
\label{fig9}
\end{figure}

\begin{figure}
\includegraphics[width=9cm,height=9cm]{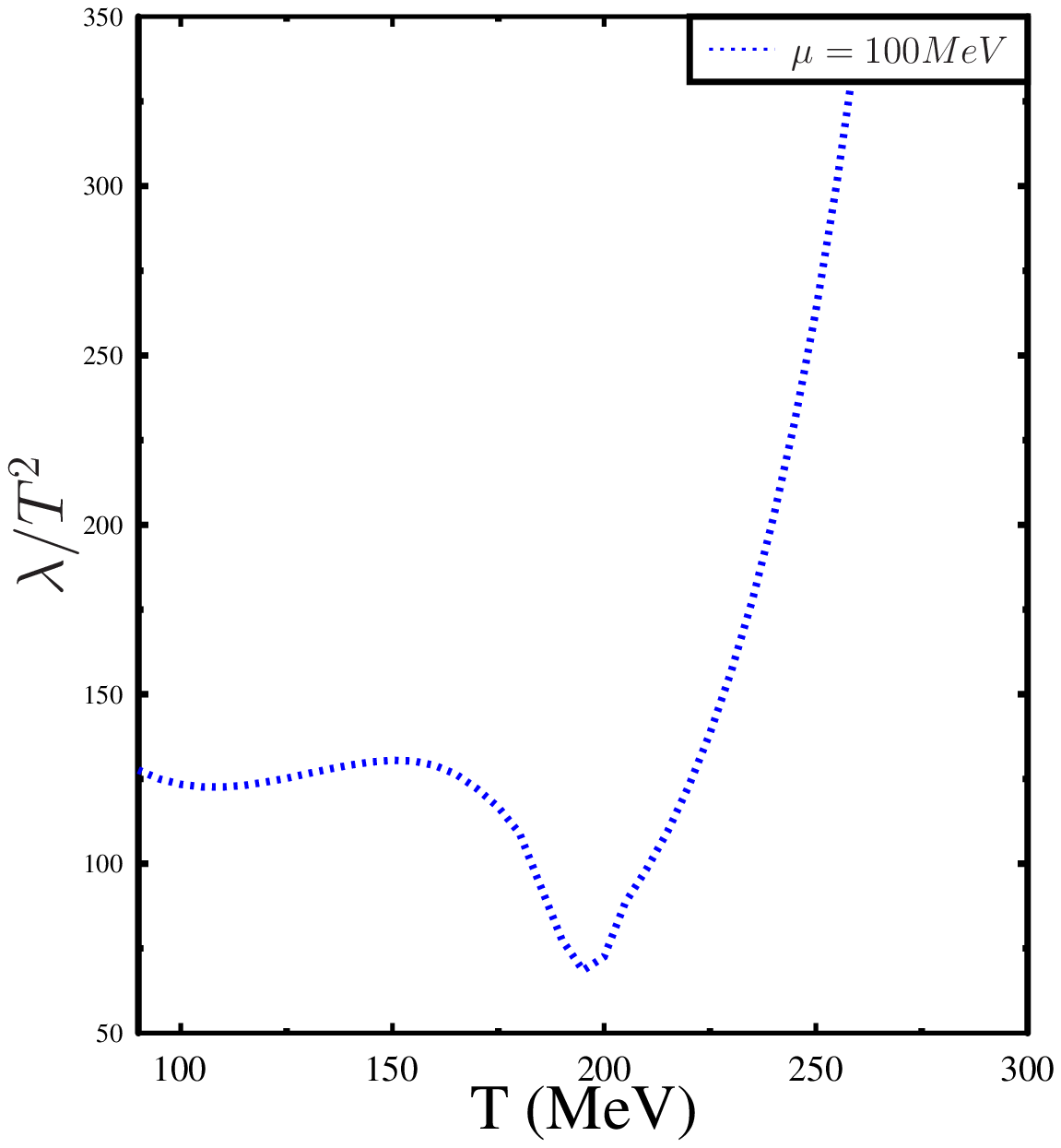}
\caption{Thermal conductivity($\lambda$) in units of $T^2$ for $\mu=100$ MeV. } 
\label{fig10}
\end{figure}

At zero baryon density, bulk viscosity depends quadratically upon the violation of conformality measures
 $C_1=1-3v_n^2$ and $C_2=M^2-TM\frac{dM}{dT}$ \cite{purnendu}. The behaviors of these two
parameters are plotted as a function of temperature in Fig. 9. For comparison, we have also plotted
the corresponding quantities at nonzero $\mu$. Both these measures peak where
the trace anomaly is maximal as shown in Fig. (2b)
As may be observed, $\zeta/s$ is largest when the violation of conformality is large. At finite baryon density, $(1-v_n^2)$ does not vanish,
nor does the factor $(\partial P/\partial n)_\epsilon$ as a result of which the ratio $\zeta/s$ does not vanish, unlike the  $\mu=0$ case.
The behavior of bulk viscosity is similar qualitatively to that of linear sigma model of Ref.\cite{purnendu}. 
Our results regarding the ratio $\eta/s$ qualitatively look similar as compared to that of Ref.\cite{sasakinjl}. However,
the bulk viscosity to entropy ratio look different as we have implemented the Landau-Lifshitz matching conditions explicitly,
leading to a different expression for $\zeta$.  Apart from this, while estimating the average relaxation time we have used the transition rate
calculated in a covariant manner similar to Ref. \cite{klevanskynpa} whereas Ref. \cite{sasakinjl} uses a probability distribution
to calculate the thermal-averaged cross section similar to \cite{klevansky}.

 Finally, in Fig. 10 we have plotted thermal conductivity of 
quark matter at $\mu=100 MeV$ in units of $T^2$. 
Let us note that thermal conduction, which involves the relative 
flow of energy and baryon number, vanishes at zero baryon density. In fact,
$\lambda$ diverges as $\lambda\sim 1/n^2$ as may be observed in Eq.(\ref{lambda2}). Such a divergence, 
however, is inconsequential because, e.g., in the dissipative current as in Eq.(\ref{deltajmu}), it enters
as $\lambda n^2$ \cite{hosoya,gavin} and the heat conduction vanishes
for $\mu=0$ \cite{daniel}.  We have therefore shown the results for thermal conductivity 
for nonvanishing $\mu$ arising from quark scatterings. 
 As may be noted,
the ratio $\lambda/T^2$ shows a nonmonotonic behavior with a minimum at
the critical temperature. The origin of this  again is related to the minimum of the relaxation time at the critical temperature.
The present behavior is in contrast to the same obtained in Ref. \cite{marty}, where, the same ratio shows a monotonically decreasing function
of temperature. The behavior of $\lambda/T^2$ was also studied in Ref.\cite{matiello}, where, the ratio showed an increasing
behavior with temperature with, however, a slower rise with temperature as compared to the results shown in Fig. 6.
 The reason for a faster rise of $\lambda$ with temperature beyond $T_{mott}$ is two fold. Firstly, 
the prefactor in Eq.{\ref {lambda2}}, $(w/nT)^2$,  varies at $T^2$, because, $w$ rises as $T^4$, while $n$ varies as $\mu T^2$ in the
 massless limit for small chemical potential. In addition, 
at large temperature, the integral itself rises as $T^3$
apart from the temperature dependence of relaxation time, which, again 
is an increasing function of
 temperature beyond $T_{mott}$.
Within the Green-Kubo approach, thermal conductivity was estimated for two flavors using NJL model\cite{fukutome} as well as  in Ref.\cite{nam} 
within the instanton liquid model where however the thermal conductivity saturates beyond T=150 MeV in contrast to the present result. 

 We may remark here however that, for situations, where, e.g. the pion number is conserved, particularly at
low temperatures, heat conductivity can be sustained by pions which themselves have zero baryon number.
This is due to the fact that it is possible to have pion scattering in a 
medium where to number of pions are conserved. This has been the basis of estimating thermal conductivity
at zero baryon density \cite{nicola,sourav,sabyath} using chiral perturbation
theory.  We may note here that, since pions are not the elementary
 degree of freedom in the NJL model,
there is no elementary pion-pion interaction within the model. However,
it is possible to construct effective meson-meson interaction
using the NJL model considering leading-order diagrams in a $1/N_c$ expansion
similar to Refs.\cite{quack,heckmann}. In such an approach, the pions are coupled through a quark loop as well as exchange of
 sigma mesons which couple to pions through quark triangles. Such an approach, in principle, can be made to estimate 
the relaxation time involving pion scattering and hence its contribution to
transport coefficients.
This is, however, beyond the scope
of the present investigation and will be reported elsewhere.

 \section{summary}
We have attempted here to compute the transport coefficient in NJL model. The approach uses solving the Boltzmann kinetic equation
within relaxation time approximation. To estimate the relaxation time we have considered the  quark-antiquark two body scatterrings 
through exchange of pion and sigma resonances. Since the meson masses are minimal at the transition temperatures beyond which
they are degenerate and increase linearly with temperature, the meson propagator occurring in the transition
amplitude lead to a large contribution to the cross section for the quark- antiquark scattering. This eventually 
leads to a smaller relaxation time which, in turn, leads to a 
minimum in the temperature dependence of the relaxation time. While computing the averaged relaxation time we have 
performed the procedure in a manifestly covariant manner as in Ref \cite{klevanskynpa}, rather than multiplying an ad hoc 
probability function to estimate the thermal-averaged cross section \cite{klevansky}.
We have used the expressions for the transport coefficients that are manifestly positive definite as they should be. The expression
for shear viscosity only depends on the relaxation time and the distribution functions. However,  the expressions
for both the coefficients of bulk viscosity and thermal conductivity involve equation of state. The expressions for 
the transport coefficients are direct generalization of their counterparts at zero chemical potential \cite{albright}. All  three
transport coefficients are minimal at the Mott temperature.

For the estimation of the relaxation time we have only included 
 two-body scatterings. 
One can generalize this to include decay processes involving the mesons
decaying to a pair of quarks and antiquarks\cite{lang,ghoshkrein}. We have investigated here the temperature dependence of the 
transport coefficients
in relation to the chiral transition in quark matter. It would be interesting to study the interplay of chiral and deconfinement transition
using a Polyakov loop NJL model.  Some of these calculations are in progress and will be reported elsewhere.

\acknowledgements
H.M. would like to acknowledge many discussions with P. Chakravarty, D. Voskresensky, M. Albright, S. Ghosh and  R. Gavai. He would also like
to thank the organizers of the  CNT-QGP workshop 2015  at VECC, Kolkata, and the Workshop on High Energy Physics Phenomenology (WHEPP)2015,
IIT Kanpur where this work was presented.

\def\berera{A. Bstero-Gil, A. Berera and R. Ramos, JCAP1107, 030 (2011).}
\def\heinzrev{U. Heinz and R. Snellings, Annu. Rev. Nucl. Part. Sci. 63, 123-151, 2013}
\def\pethick{H. Heiselberg and C. Pethick,{\PRD{48}{2916}{1993}}.}
\def\exptstar{J. Adams,{\em et al}(STAR),{\NPA{757}{102}{2005}}.}

\def\karschkharzeev{F. Karsch, D. Kharzeev, and K. Tuchin, Phys. Lett. B
663, 217 (2008).}
\def\kharzeevtuchin{D. Kharzeev, and K. Tuchin,JHEP 0808,031, (2008).}
\def\nicolaprl{	
D. Fernandez-Fraile, A. Gomez Nicola, {\PRL{102}{121601}{2009}}.}
\def\joglekar{J.C. Collins, A. Duncan, S.D. Joglekar, Phys. Rev. D 16, 
438 (1977).}
\def\blaschke{J. Jankowski, D. Blaschke, M.Spalinski, Phys.Rev.D 87, 105018^M
(2013). }
\def\gorenstein{M. Gorenstein, M. Hauer, O. Moroz, Phys.Rev.C 77,024911 (2008)}
\def\bugaev{K. Bugaev et al, Eur.Phys.J. A 49, 30 (2013)}
\def\cpsingh{S.K. Tiwari, P.K. Srivastava, C.P. Singh, Phys.Rev. C 85,
014908 (2012) }
\def\chen{J.-W. Chen, Y-H. Li, Y.-F. Liu, and E. Nakano, Phys. Rev. D 76,
114011 (2007)}
\def\chennakano{J.-W. Chen, and E. Nakano, Phys. Lett. B 647, 371 (2007)}
\def\itakura{K. Itakura, O. Morimatsu, and H. Otomo, Phys. Rev. D 77, 014014
(2008)}
\def\cleymans{J. Cleymans, H. Oeschler, K. Redlich, and S. Wheaton, Phys.
Rev. C 73, 034905 (2006)}
\def\worku{J. Cleymans and D. Worku, Mod. Phys. Lett. A26,1197,(2011).}
\def\guptagod{S. Chatterjee, R. M. Godbole and S. Gupta, {\PRC{81}{044907}{2010}}.}
\def\Noronha{Noronha-Hostler J, Noronha J and Greiner C 2012 Phys. Rev. C 86 024913}
\def\tanmoy{A. Bazavov {\it etal}, e-print:arXiv:1407.6387.}
\def\cavitation{K. Rajagopal and N. Trupuraneni, JHEP1003, 018(2010);
 J. Bhatt, H. Mishra and V. Sreekanth, JHEP 1011, 106,(2010);{\it ibid} Phys. Lett. B704, 486 (2011); {\it ibid} Nucl. Phys. A875, 181(2012).}
\def\borsonyi{S. Borsonyi{\it etal}, JHEP1011, 077 (2010).}
\def\borsonyimu{S. Borsonyi{\it etal}, JHEP1208, 053 (2012).}
\def\dobadoch{A. Dobado,F.J.Llane-Estrada amd J. Torres Rincon, 
{\PLB{702}{43}{2011}}.}
\def\dobadoshear{A. Dobado,F.J.Llane-Estrada amd J. Torres Rincon, 
{\PRD{79}{055207}{2009}}.}
\def\sasakiqp{C. Sasaki and K.Redlich,{\PRC{79}{055207}{2009}}.}
\def\sasakinjl{C. Sasaki and K.Redlich,{\NPA{832}{62}{2010}}.}
\def\ellislet{I.A. Shushpanov, J. Kapusta and P.J. Ellis,{\PRC{59}{2931}{1999}}
; P.J. Ellis, J.I. Kapusta, H.-B. Tang,{\PLB{443}{63}{1998}}.}
\def\prakashwiranata{Anton Wiranata and Madappa Prakash,
{\PRC{85},{054908}{2012}}.}
\def\purnendu{P. Chakravarti and J.I. Kapusta {\PRC{83}{014906}{2011}}.}
\def\greco{S.Plumari,A. Paglisi,F. Scardina and V. Greco,{\PRC{83}{054902}{2012}a.}}
\def\bes{H. Caines, arXiv:0906.0305 [nucl-ex], 2009.}
\def\greinerprl{J. Noronha-Hostler,J. Noronha and C. Greiner,
{\PRL{103}{172302}{2009}}.}
\def\greinerprc{J. Noronha-Hostler,J. Noronha and C. Greiner
, {\PRC{86}{024913}{2012}}.}
\def\igorgreiner{J. Noronha-Hostler, C. Greiner and I. Shovkovy,
, {\PRL{100}{252301}{2008}}.}
\def\majumdermueller{A. Majumder and B. Mueller, {\PRL{105}{252002}{2010}}.}
\def\leonidov{ A. V. Leonidov and P. V. Ruuskanen, {\EPJC{4}{519}{1998}}.}
\def\cbm{ B. Friman, C.H. Ohne, J. Knoll, S. Leupold, J. Randrup, R. Rapp, P. Senger (Eds.), Lect. Notes Phys., vol. 814,
2011.}
\def\nica {A.N. Sissakian, A.S. Sorin, J. Phys. G 36 (2009) 064069.}
\def\nakano{J.W. Chen,Y.H. Li, Y.F. Liu and E. Nakano,
 {\PRD{76}{114011}{2007}}.}
\def\wang{M.Wang,Y. Jiang, B. Wang, W. Sun and H. Zong, Mod. Phys. lett.
{\bf A76}, 1797,(2011).}
\def\agasian{N.O. Agasian, JETP Lett. 95, 171, (2012), arXiv:1109.5849.}
\def\Hagedorn{R. Hagedorn and J. Rafelski,{\PLB{97}{136}{1980}}.}
\def\kapustaolive{J.I. Kapusta and K. A. Olive, {\NPA{408}{478}{1983}}.}
\def\hrgexp{P. Braunmunzinger, J. Stachel, J.P. Wessels and N. Xu,
{\PLB{365}{1}{1996}}; G.D. Yen and M.I. Gorenstein, {\PRC{59}{2788}{1999}};
F. Becattini, J. Cleymans, A. Keranen, E. suhonen and K. Redlich, 
{\PRC{64}{024901}{2001}}.}
\def\rischkegorenstein{.D.H. Rischke, M.I. Gorenstein, H. Stoecker and
W. Greiner, Z.Phys. C {\bf 51}, 485 (1991).}
\def\hmnjl{Amruta Mishra and Hiranmaya Mishra, {\PRD{74}{054024}{2006}}.}
\def\pdgb{C. Amseler {\it et al}, {\PLB{667}{1}{2008}}.}
\def\shuryak{E.V. Shuryak, Yad. Fiz. {\bf 16},395, (1972).}
\def\leupold{S. Leupold, J. Phys. G{\bf32},2199,(2006)}
\def\peter{A. Andronic, P. Braun-Munzinger , J. Stachel and M. Winn,
{\PLB{718}{80}{2012}}}
\def\blum{M. Blum, B. Kamfer, R. Schluze, D. Seipt and U. Heinz,{\PRC{76}{034901}{2007}}.}
\def\jaminplb{M. Jamin, {\PLB{538}{71}{2002}}.}
\def\ghosh{Sabyasachi Ghosh, {\PRC{90}{025202}{2014}}; International Journal Of Modern Physics {\bf A29}, 145005,2014.}
\def\csernai{L.P. Csernai, J.I. Kapusta and L.D. McLerran,{\PRL{97}{152303}{2006}}.}
\def\hagedorn{R. Hagedorn, Nuovo Cim. Suppl. 3,147 (1965); Nuovo Sim. A56,1027 (1968).}
\def\torieri{G. Torrieri and I. Mishustin,{\PRC{77}{034903}{2008}}.}
\def\fernandez{D. Fernandiz-Fraile and A.G. Nicola,{\PRL{102}{121601}{2009}}.}
\def\caron{S.Caron,{\PRD{79}{125009}{2009}}.}
\def\latticemeyer{H.B. Meyer,{\PRL{100}{162001}{2008}}.}
\def\romatschke{P.Romatscke and D.T. Son,{\PRD{80}{065021}{2009}}.}
\def\moore{G.D. Mooore and O. Sarem, J. High Energy Phys. JHEP0809(2008)015.}
\def\dobado{A.Dobado and J. M. Torres-Rincon {\PRD{86}{074021}{2012}}.}
\def\monai{A. Monnai and T. Hirano, Phys. Rev. C 80, 054906 (2009).}
\def\kodama{G. S. Denicol, T. Kodama, T. Koide, and P. Mota, Phys. Rev. C 80, 064901 (2009).}
\def\heinz{H. Song and U. Heinz, Phys. Rev. C 81, 024905 (2010).}
\def\bozek{P. Bozek, Phys. Rev. C 81, 034909 (2010).}
\def\schaferdus{K. Dusling and T. Schafer,  ̈ Phys. Rev. C 85, 044909 (2012).}
\def\noronhahydro{J. Noronha-Hostler, G. S. Denicol, J. Noronha, R. P. G. Andrade,
and F. Grassi, Phys. Rev. C 88, 044916 (2013); J. Noronha-Hostler, J. Noronha and F. Grassi,
Phys. Rev. C 90, 034907 (2014)}
\def\rosegale{J. B. Rose, J. F. Paquet, G. S. Denicol, M. Luzum, B. Schenke, S. Jeon and C. Gale, 
{\NPA{931}{926}{2014}}.}
\def\bassprl{ N.~Demir and S.~A.~Bass,
Phys. Rev. Lett. {\bf 102}, 172302 (2009).}
\def\phsdbrat{V.~Ozvenchuk, O.~Linnyk, M.~I.~Gorenstein, E.~L.~Bratkovskaya and W.~Cassing, Phys. Rev. C {\bf 87},  064903 (2013).}
\def\florkowski{W. Broniowski and W. Florkowski, {\PLB{673}{142}{2009}}.}
\def\albright{M. Albright and J.I. Kapusta,{\PRC{93}{014903}{2016}}.}
\def\hirano{P. Romatschke and U. Romatschke, Phys. Rev. Lett.{\bf 99},172301, (2007); T. Hirano and
 M. Gyulassy, Nucl. Phys. {\bf A 769}, 71, (2006).}
\def\daniel{P. Danielewicz, M. Gyulassy, {\PRD{31}{53}{1985}}.}
\def\kss{P. Kovtun, D.T. Son and A.O. Starinets, Phys. Rev. Lett.{\bf 94},
 111601, (2005).}
\def\schenke{C.Gale, S. Jeon and B. Schenke, International Journal of Modern Physics A {\bf 28}, 134011,(2013).}
\def\denicolhydro{G.S. Denicol, H. Niemi, E. Molnar and D.H. Rischke,{\PRD{85}{114047}{2012}}.}
\def\denicolpre{M. Greif, F. Reining, I. Bouras , G.S. Denicol, Z. Xu and  C. Greiner, Phys.Rev. {\bf E87} ,033019(2013).}
\def\denicolheat{G.S. Denicol, H. Niemi, I. Bouras E. Molnar , Z. Xu , D.H. Rischke, C. Greiner ,{\PRD{89}{074005}{2014}}.}
\def\rincon{J.I. Kapusta and J.M. Torres-Rincon,{\PRC{86}{054911}{2012}}.}
\def\rinconprd{J.I. Kapusta and J.M. Torres-Rincon,{\PRC{86}{054911}{2012}}.}
\def\ghoshthermal{Sabyasachi Ghosh, International Journal of Modern Physics {\bf E24},1550058,2015. }
\def\kubo{R. Kubo,J. Phys. Soc. Jpn. {\bf 12},570,(1957).}
\def\klevansky{P. Zhuang,J. Hufner, S.P. Klevansky and L. Neise,{\PRD{51}{3728}{1995}}. } 
\def\klevanskynpa{P. Rehberg, S.P. Klevansky and ,J. Hufner,{\NPA{608}{356}{1996}}. } 
\def\transqcd{P. Arnold,G.D. Moore and L.G. Yaffe,JHEP, 11, 2000, 001; ibid, JHEP 01 (2003) 030; ibid, JHEP 05 (2003) 051}
\def\quasip{M. Bluhm, B. Kamfer and K. Redlich,{\PRC{79}{055207}{2009}}.}
\def\blum{M. Bluhm, B. Kamfer and K. Redlich,{\PRC{84}{025201}{2011}}.}
\def\marty{R. Marty, E. Bratkovskaya, W. Cassing, J. Aichelin and H . Berrehrah, {\PRC{88}{045204}{2013}}.}
\def\voskresenskynpaa{A.S. Khvorostukhin, V.D. Toneev and D.N. Voskresensky,{\NPA{915}{158}{2013}}.}
\def\voskresensky{A.S. Khvorostukhin, V.D. Toneev and D.N. Voskresensky,
{\NPA{845}{106}{2010}}.}
\def\gavin{Sean Gavin,{\NPA{435}{826}{1985}}.}
\def\hosoya{A. Hosoya and K. Kajantie ,{\NPB{250}{666}{1985}}.}
\def\degroot{ S.R. deGroot, W.A. van Leeuwen and Ch. G. van Weert,{\it Relativistic Kinetic Theory: Principles and Applications( North-Holland, Amsterdam, 1980)}.}
\def\lang{R. Lang, N. Kaiser, W. Weise, Eur. Phys. {\bf A48}, 109, 2012.}
\def\langweise{R. Lang, N. Kaiser, W. Weise, Eur. Phys. {\bf A50 }, 63, 2014.}
\def\ghoshkrein{Sabyasachi Ghosh, Thiago C. Peixoto, Victor Roy, Fernando E. Serna, Gastão Krein e-Print: arXiv:1507.08798 [nucl-th], (2015).}
\def\mpiexpt{K. Hagiwara {\em et al},{\PRD{66}{010001}{2002}}.}
\def\fpiexpt{B. Holostein,{\PLB{244}{83}{1990}}.}
\def\condsum{H.G. Dosch and S. Narrison,{\PLB{417}{173}{1998}}.}
\def\condlat{L. Giusti,F. Rapuano, M. Talevi and A. Vladikas,{\NPB{538}{249}{1999}}.}
\def\matiello{S. Matiello, arXiv:1210.1038[hep-ph].}
\def\nicola{ D. Fernandiz-Fraile and A. Gomez Nicola, {\EPJC{62}{37}{2009}}.}
\def\nam{ S.  Nam, Mod. Phys. Lett. A 30, 1550054,2015.}
\def\fukutome{ M. Iwasaki and T. Fukutome, J. Phys. G36, 115012, 2009.}
\def\sourav{ S. Mitra and S. Sarkar,{\PRD{89}{054013}{2014}}.}
\def\sabyath{S. Ghosh, Int.J.Mod.Phys. E24 (2015) 07, 1550058}
\def\prakash{ M. Prakash, M. Prakash, , R. Venugopalan and G. Welke,{\PR{227}{321}{1993}}.}
\def\buballarev{M. Buballa,{\PR{407}{205}{2005}}.}
\def\heckmann{K. Heckmann, M. Buballa and J. Wambach ,{\EPJA{48}{142}{2012}}}.
\def\quack{E. Quack, P. Zhuang, Y. Kalinovsky, S.P. Klevansky and
J. Hufner,{\PLB{348}{1}{1995}}}.
\def\zhuang{P. Zhuang,J. Hufner, S.P. Klevansky {\NPA{576}{525}{1994}}. } 
\def\gondolo{J. Edsjo, and P. Gondolo, {\PRD{56}{1879}{1997}}. } 
\def\goity{J. I. Goity, and H. Leutwyler, {\PLB{228}{517}{1989}}. }

\end{document}